\documentclass[aps,reprint,notitlepage,prx,nofootinbib]{revtex4-2}
\usepackage{amsmath,amsfonts,amssymb,mathrsfs,bm,MnSymbol}
\usepackage{marginnote}
\usepackage[colorlinks]{hyperref}
\usepackage{censor}
\newlength\nextcharwidth
\makeatletter
\renewcommand\@cenword[1]{%
  \setlength{\nextcharwidth}{\widthof{#1}}%
  \censorrule{\nextcharwidth}%
  \kern -\nextcharwidth%
  #1}
\makeatother

\hypersetup{citecolor=blue}
\numberwithin{equation}{section}
\usepackage{graphicx}
\usepackage{xcolor}
\usepackage[normalem]{ulem}
\usepackage{cancel}

\renewcommand{\vec}[1]{\boldsymbol{#1}}
\newcommand{\tens}[1]{\mathbf{#1}}

\newcommand{\kbT}{k_{\scriptscriptstyle\textrm{B}}T}

\newcommand{\mean}[1]{{\langle #1 \rangle}}
\newcommand{\uu}{\hat{\vec{u}}}
\newcommand{\uvec}[1]{\hat{\vec{#1}}}
\newcommand{\Rhat}{\hat{\vec{R}}}

\begin{document}

\counterwithout{equation}{section}
\renewcommand{\theequation}{\arabic{equation}}

\title{A Smoluchowski equation for a sheared suspension of frictionally interacting rods } 

\author{Genevieve Qui\~nones}
\author{Peter D. Olmsted}
\email[Corresponding author: ]{pdo7@georgetown.edu}
\affiliation{Department of Physics and Institute for Soft Matter Synthesis and Metrology, Georgetown University, 3700 O St NW, Washington DC 20057, USA}

\date{\today}
\begin{abstract}
In this work we develop constitutive equations for a dense, sheared suspension of frictionally interacting rods by applying Onsager's variational method as formulated by Doi. We treat both solid friction, of the Amontons-Coulomb form; and lubricated friction, which scales with relative tangential velocity at the contact point. Dissipation functions in terms of the rod angular velocity are derived via a mean field approach for each form of friction, and from these, a Rayleighian for dense suspensions of rigid rods under shear constructed. Derivatives of this Rayleighian with respect to rod angular velocity and velocity gradient give a Smoluchowski equation and stress tensor, respectively. We show that these are representable as perturbations to Doi's model for a sheared liquid crystal. We also suggest a form for the average number of contacts between rods as a function of volume fraction, aspect ratio, and nematic order parameter, generalizing Philipse's random contact equation for disordered packings.
\end{abstract}

\pacs{}

\maketitle 

\section{Introduction}

The effects of flow on suspensions of rigid rod-like particles has been studied for more than a century \cite{bjorkman}. Rods are typically considered to interact through a combination of hydrodynamic forces and potential energies such as electrostatics  van der Waals interactions \cite{odijk,dhontbriels}. It has been suggested that particles in dense suspensions under shear come into solid contact, breaking lubrication films \cite{fernandez,lootens} to then  interact through frictional forces. The presence of friction in rods undergoing shear flow prompts several questions analogous to those studied for spheres: how does friction affect the orientational state and the stress? Can these interactions lead to discontinuous shear thickening and the onset of jamming, in the fashion of the Wyart and Cates \cite{wyartcates} model? How many contacts are there per rod in a flowing system, compared to that found in a disordered packing \cite{randomcontact}, and how does this contact number depend on the degree of orientational order? These questions motivate us to develop a theoretical description of a suspension of rigid rods under flow, to explicitly account for contacts between rods.

The dilute and semi-dilute regimes of suspensions of rigid rods are characterized by a volume fraction $\phi$ and aspect ratio $L/D$ satisfying 
\begin{subequations} \label{dilu-to-semi}
\begin{align}
  \phi \frac{L^2}{D^2} &\ll 1& &\textrm{(dilute)}  \\
   \frac{D}{L} \ll \phi \frac{L}{D} & < 1 &&\textrm{(semi-dilute),}
\end{align}
\end{subequations}
where $D$ is the rod diameter and $L$ rod length \footnote{See ~\hyperref[table:nomen]{Nomenclature} for general notation.}. Various theories have successfully modeled orientational dynamics under flow for these regimes (\citet{jeff}, \citet{advani}, \citet{doimain}). The dimensionless Peclet number $\textrm{Pe}$ weighs the effects of hydrodynamic and Brownian forces in rod dynamics, $\text{Pe} \equiv \dot{\gamma}/D_{r}$, where $\dot{\gamma}$ is the rate of shear and $D_{r}$ the rotational diffusion constant. High $\text{Pe}$ is often referred to as ``non-Brownian".

Many studies begin with  Jeffery's \cite{jeff} equation of motion $\dot{\vec{u}}_{\text{Jeffery}}$ for the rotation of a single ellipsoid suspended in a Newtonian fluid in a flow field, and insert this into the continuity equation for the probability distribution of rod orientations (\ref{psi-gen}) to develop a Smoluchowski equation that accounts for the effect of rod-rod interaction.
Doi \cite{doimain} addressed the general $\text{Pe}$ case, and used  an expression for the excluded volume of rods as a function of volume fraction, aspect ratio, and nematic order $S$ to calculate a dynamical equation for the $\tens{Q}$ tensor, which is defined as an orientational average of particle vectors:
\begin{equation} \label{q-tens-def}
\begin{split}
Q_{\alpha \beta} & = \left\langle \hat{u}_{\alpha} \hat{u}_{\beta} - \frac{\delta_{\alpha \beta}}{3} \right\rangle \\
& = S \left(\hat{n}\hat{n} - \frac{\delta_{\alpha \beta}}{3} \right).
\end{split}
\end{equation}
Here, $\uu$ refers to the orientation of a single particle, $\hat{\vec{n}}$ is the director, and the second line follows for a $uniaxial$ $\tens{Q}$. \citet{advani} applied this method to the the non-Brownian case and included a phenomenological term proportional to the rotary diffusivity (attributed to random shear-induced rod-rod collisions) to find a dynamical equation for $\tens{Q}$ similar to Doi's. \citet{folgartucker} suggested a similar form, with a rotary diffusion constant that depended on the magnitude of the symmetric velocity gradient tensor, $|\dot{\gamma}|$. \citet{phan1991} proposed a constitutive equation for spatially homogeneous fiber suspensions that expanded Ericksen's transversely isotropic fluid model \cite{ericksen1960} to the semi-dilute regime and successfully modeled specific viscosity varying as volume fraction cubed. \citet{dinharmstrong} presented a constitutive model for the high Peclet regime that included a contribution to the stress from a given orientational state; this model was used by \cite{tsengmelt} to calculate viscosity overshoots in fiber filled polymer melts under shear.

\begin{figure}
\centering\includegraphics[width=0.495\textwidth]{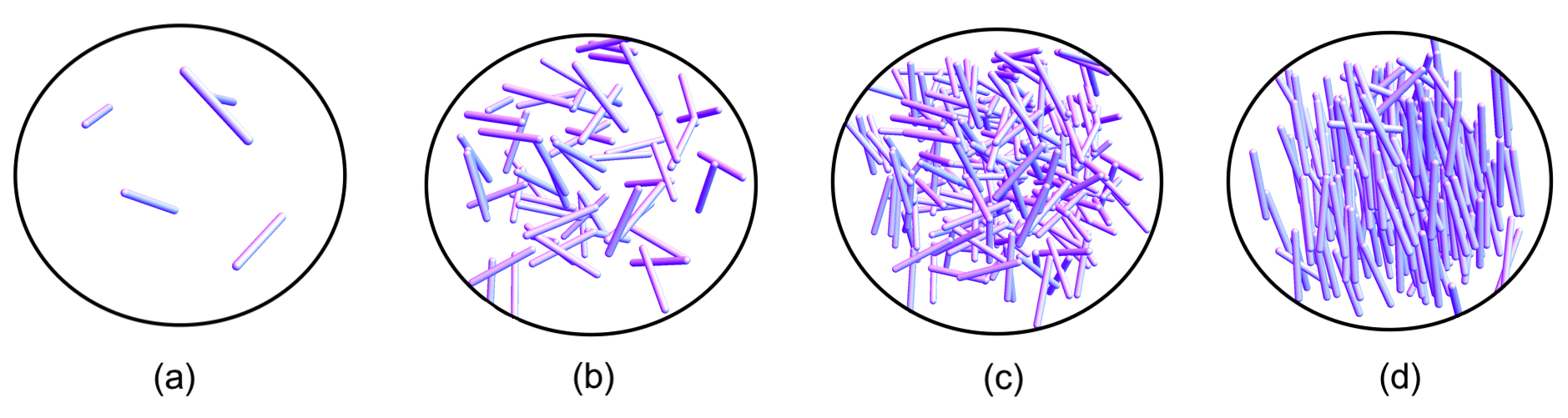}
\caption{Various concentration regimes of rod suspensions: a) dilute, b) semi-dilute, c) concentrated, for $S = 0$, d) concentrated, for $S > 0$.}\label{fig:phi-change}
\end{figure}

In the concentrated regime,
\begin{equation}
    \phi \frac{L}{D} > 1
\end{equation}
and rod-rod interactions dominate \cite{doi1978}. For solutions with $\phi L/D \gtrsim 3$, the isotropic-nematic transition occurs in Brownian suspensions \cite{doionsager}, leading to non-zero orientational order $S > 0$ (see Fig.~\ref{fig:phi-change}). Some theories \cite{advani,doimain,dinharmstrong} that successfully model the semi-dilute regime and capture orientational dynamics such as flow alignment, tumbling, and kayaking \cite{larsonottinger}, can extend to larger $\phi L/D$ and also model orientational properties of dense, sheared rods. However, they can’t explain certain rheological behaviors observed in concentrated solutions of rodlike particles. These behaviors include discontinuous shear thickening (DST), where the viscosity discontinuously jumps in value with an increase in shear rate, shear jamming, a state induced in a dense system which at rest is  below the jamming concentration $\phi_{j}$ \cite{panreview}, (\citet{goyaljaeger}, \citet{tapia}, \citet{egres}) and vorticity tilting (\citet{rathee}), the observed tilt of the director into the vorticity direction (out of the shear plane) in sheared rod suspensions. \citet{pipes1994} considered a ``hyperconcentrated” regime, with highly aligned fibers of aspect ratio $> 100$, but only predicted shear thinning and did not address any other rheological effects. The failure of these models is due in part to their implicit assumptions, such as uncorrelated rod orientations, breaking down due to increased  rod interactions at higher volume fraction and aspect ratio; but also due to the neglect of \textit{frictional} inter-particle contacts \cite{singh2023hidden}.

Recent experimental and computational work has suggested that such unexplained rheological behaviors are indeed due to frictional contacts (\citet{setomorris}, \citet{royer}). For suspensions of spheres, Wyart and Cates \cite{wyartcates} established a theory that relates the fraction of frictional contacts to DST and shear jamming. In concentrated suspensions of large aspect ratio rods, a given rod may have many contacts and remain below the Maxwell threshold \cite{maxwell1864}, so frictional contacts may play an even larger role in the pre-jam dynamics of rods than in spheres. Nematic order $S$ is also expected to affect the average number of contacts per particle, which is larger for a greater average amount of excluded volume per interacting pair \cite{randomcontact}. A particle less aligned with its neighbors (\emph{i.e.}, lower $S$) has a larger average excluded volume and so more contacts. Thus nematic order, a variable not present in spherical particle suspensions, is relevant to the dynamics of dense rod suspensions and must also be accounted for.

Frictional contacts have been included for rods by  \citet{sandstrom} and  \citet{tollmanson} in two dimensional ``planar" fiber models; they obtained  expressions for the stress tensor and  angular velocity for fibers interacting with frictional forces linear in the relative tangential velocity at contact (hereafter referred to as ``boundary lubricated friction"). Similarly, for the 3D case \citet{djalili} explored the effect on suspension stress from rod contacts that interacted with a boundary lubricated frictional force, though they did not address the effects on equations of motion. \citet{bounoua} studied both sliding Coulomb friction and boundary lubricated frictional interactions between rods in three dimensions. They started from Jeffery's equation and constructed an additional torque from frictional contacts between rods. They assumed the orientational dynamics that determined relative rod-rod velocity (and thus frictional forces) was exactly Jeffery's equation, \emph{i.e.} simply the motion in the dilute case. As in \cite{sandstrom} (see also Appendix \ref{appendix:fiber-torque}), they found this additional torque to be zero, arguing that the torque due to contacts vanished, based on the assumption that contacts are distributed symmetrically about a rod's center.

Despite the null result of \citet{bounoua}, an important point stands: frictional contacts are a source of dissipation in dense suspensions, and should lead to an accompanying torque. Doi \cite{doionsager} showed how to implement Onsager's dissipation principle to re-derive his model for flowing liquid crystals \cite{doionsager}, by minimizing the total dissipation (known as the ``Rayleighian" \cite{rayleigh1896}) in a suspension of rods under shear, which leads  
to an equation of motion for $\dot{\vec{u}}$ from the corresponding balance of torques, each contribution arising from a separate source of dissipation. This is in contrast to previous  methodologies \cite{bounoua,sandstrom} that relied on assuming  an equation of motion  $\dot{\vec{u}}$ appropriate for  dilute rods. We show here that by following the framework of Doi and including the contribution to total dissipation due to frictional sliding between rods in contact, the proper frictional torque within $\dot{\vec{u}}$ can be derived. This will result in a self-consistent equation of motion, as this torque is naturally a function of $\dot{\vec{u}}$. 

First, we outline the general method of constructing a Rayleighian and calculating a corresponding Smoluchowski equation from it in §~\ref{sec:general-derivation}. Then we propose a mean field average based on the typical distribution of contacts on a given rod in §~\ref{sec:contact-field}, which we use in §~\ref{sec:Smol} to derive an addition to the canonical Rayleighian for dense rod suspensions that accounts for the dissipation from frictional contacts. From the Rayleighian we calculate a new Smoluchowski equation in §~\ref{subsec:new-smol}. Finally, we use the new Smoluchowski equation to derive a dynamical equation for the $\tens{Q}$ tensor in §~\ref{subsec:new-q-tens}, and calculate a stress tensor from the Rayleighian in §~\ref{subsec:new-stress}.

\section{Rayleighian Derivation of the Smoluchowski Equation} \label{sec:general-derivation}
\subsection{General Approach}

We follow \citet{doibook} to derive a Smoluchowski equation for a suspension of rods. First, a Rayleighian ${\cal R}(\vec{\omega})$ is written as the sum of the rate of change of free energy of the rods $\dot{A}$ and the dissipation due to rod angular velocity $\vec{\omega} = \hat{\vec{u}} \times \dot{\vec{u}}$, expressed via a dissipation function $\Phi$:
\begin{equation} \label{rayl-def}
     \mathcal{R} = \Phi(\vec{\omega}) + \dot{A}.
\end{equation}

For example, Doi \cite{doibook} considered a simple drag constant $\zeta_r$,  which leads to a drag torque on the $i$th rod of
\begin{equation} \label{drag}
    \vec{\Gamma}^i_{\text{drag}} = -\zeta_{r} \vec{\omega}^i. 
\end{equation}
 By integrating to get power $P^i=-\int\vec{\Gamma}^i_{\textrm{drag}}\cdot d\vec{\omega}^i$ we find the dissipative contribution to the Rayleighian: 
\begin{subequations} \label{rayl-drag}
\begin{align}
\Phi^i_{\text{drag}} &= -\int\vec{\Gamma}^i_{\textrm{drag}}\cdot d\vec{\omega}^i \\
&= \frac{1}{2}\zeta_{r} (\omega^i)^2.\label{eq:Rdiscrete}
\end{align}
\end{subequations}
We can compute the average dissipation per rod due to drag by integrating Eq.~\ref{rayl-drag} over all possible rod orientations $\hat{\vec{u}}$,
\begin{equation}
\Phi_{\text{drag}} = \frac{1}{2}\int\zeta_{r} \omega^{2} \psi(\hat{\vec{u}}) \,\text{d} \hat{\vec{u}}, \label{eq:Ravg}
\end{equation}
where we have introduced the distribution function $\psi(\hat{\vec{u}})$ of rod orientations.

The Rayleighian is differentiated with respect to $\vec{\omega}$ to obtain a net torque
\begin{equation}
\vec{\Gamma}(\vec{\omega}) = \frac{\delta R}{\delta\vec{\omega}}. \label{eq:torque}
\end{equation}
[For  a Rayleighian function (as in Eq.~\ref{eq:Rdiscrete}), $\vec{\Gamma}^i=\partial{\cal R}/\partial\omega_i$, while for a Rayleighian functional (Eq.~\ref{eq:Ravg}) we use  a variational derivative $\vec{\Gamma}=\delta{\cal R}/\delta\omega$.] The condition of zero total torque 
$\vec{\Gamma}(\vec{\omega}_{\textrm{min}})=0$ corresponds to  minimizing the Rayleighian, and the resulting
angular velocity  $\vec{\omega}_{\text{min}}$ is easily translated to the equation of motion for the rod orientation,
\begin{equation}
\frac{\partial\uu(t)}{\partial t}\equiv\vec{\omega}_{\text{min}}\times\uu.
\end{equation}

The Smoluchowski Equation for the orientational distribution function $\psi$  can be calculated by using $\vec{\omega}_{\text{min}}$  to calculate the continuity equation for probability density:
\begin{equation} 
\dot{\psi} = - \Rhat \cdot (\vec{\omega}_{\text{min}} \psi), \label{psi-gen}
\end{equation}
where $\Rhat$ is the Doi-Edwards rotational differentiation operator,
\begin{equation} \label{rot-operator}
\Rhat = \hat{\vec{u}} \times \frac{\partial}{\partial \hat{\vec{u}}}.
\end{equation}

The average free energy per particle is given by
\begin{equation} \label{freeA}
 A = \int\text{d}\hat{\vec{u}} 
 \left[\kbT\ln{\psi} +  U(\hat{\vec{u}}) \right]\,\psi,
\end{equation}
where $U$ is the inter-rod potential and $\kbT\ln{\psi}$ the so-called ``Brownian potential". The corresponding rate of change of the energy in the presence of rod rotation is \footnote{In Einstein notation this integrand is $\psi \omega_{i} \epsilon_{ijk} u_{j} \left(\frac{\partial }{\partial u_{k}}\left[\kbT\ln{\psi} + U(\hat{\vec{u}}) \right]\right)$, where $\epsilon_{ijk}$ is the Levi-Civita tensor.}
\begin{subequations} \label{fe-dissi}
\begin{align}
\dot{A} = \int \vec{\omega} \cdot \psi \Rhat\left[\kbT\ln{\psi} + U(\hat{\vec{u}}) \right]\,\text{d} \hat{\vec{u}},\label{eq:RU}
\end{align}
\end{subequations}
so the Rayleighian becomes
\begin{equation} \label{R-no-flow}
    \mathcal{R}_{0} = \tfrac{1}{2}\int{\zeta_{r} \omega^{2} \psi \text{d} \hat{\vec{u}}}  + \int \vec{\omega} \cdot \psi \Rhat\left[\kbT \ln{\psi} + U(\hat{\vec{u}})\right]
    \text{d} \hat{\vec{u}}.
\end{equation}
Upon taking the derivative $\delta \mathcal{R}_{0}/\delta \vec{\omega}$, we find the torque
\begin{equation} \label{omega0}
\boldsymbol{\Gamma}=\zeta_r\vec{\omega} + \Rhat\left[\kbT \ln{\psi} + U(\hat{\vec{u}}) \right].
\end{equation}
Note that this is the torque on the fluid due to the particles. By demanding zero net torque, we find 
\begin{equation}
    \vec{\omega}_{\textrm{min}}=-\Rhat\left[\kbT\ln{\psi} + U(\hat{\vec{u}}) \right]/\zeta_r,
\end{equation}
which together with Eq.~\ref{psi-gen} gives the Smoluchowski equation of \citet{doimain}:
\begin{equation}
\dot{\psi}= \frac{1}{\zeta_r}\Rhat\cdot\left[\psi \Rhat\left(\kbT\ln\psi + U\left(\uu\right)\right) \right].
\end{equation}
If the suspension is under flow with velocity gradient 
\begin{equation}  \label{veloc-grad-def}
(\nabla \vec{v})_{\alpha \beta} \equiv \frac{\partial v_{\alpha}}{\partial x_{\beta}},
\end{equation}
then the dissipation function due to drag becomes \footnote{Due to the dissipation of the solvent viscosity, the Rayleighian also gains a term $\eta_{s} (\nabla \vec{v} + \nabla \vec{v}^{T}):(\nabla \vec{v} + \nabla \vec{v}^{T})$, but this is irrelevant in calculating $\vec{\omega}_{\text{min}}$.}
\begin{equation} \label{shear-dissi}
    \Phi_{\text{drag}} =  \tfrac{1}{2}\int{\zeta_{r} \left[(\vec{\omega} - \vec{\omega}_{0})^{2} + \frac{1}{2} (\uu \cdot \nabla \vec{v} \cdot \uu)^2 \right] \psi \text{d} \hat{\vec{u}}},
\end{equation}
where 
\begin{equation} \label{flow-omega}
\vec{\omega}_{0} = \uu \times (\nabla \vec{v} \cdot \uu)
\end{equation}
is the angular velocity provided to a lone rod by the flow, and the second term in Eq.~\ref{shear-dissi} represents the dissipation of a rod rotating exactly with $\vec{\omega}_{0}$. Note that for the large aspect ratio limit, $\vec{\omega}_{0}$ is exactly Jeffery's equation \cite{jeff}. Upon extremizing the Rayleighian, we find
\begin{equation} \label{doi-rayl-flow-omega}
\vec{\omega}_{\text{min}}=  \vec{\omega}_{0} -\frac{1}{\zeta_{r}}\Rhat\left[\kbT\ln{\psi} + U(\hat{\vec{u}}) \right],
\end{equation}
which is the equation of motion of a rod in the presence of  both the molecular potential $U(\uu)$ and the Brownian potential $\kbT\ln\psi$. This leads to a corresponding Smoluchowski equation:
\begin{equation}\label{eq:SmolDOI}
    \dot{\psi}= \frac{1}{\zeta_r}\Rhat\cdot\left[\psi\Rhat\left(\kbT\ln\psi + U\left(\uu\right)\right)\right] -\Rhat \cdot \left( \vec{\omega}_{0} \psi \right).
\end{equation}
This expression does not take into account hydrodynamic interactions between rods. Here, we will ignore these, due to the complexity of their general inclusion. See \cite{yamane} for a treatment of lubrication interactions from a Rayleighian perspective. Although the hydrodynamic form of lubrication theory has a divergent force upon approaching contact, which prohibits solid contact, this approximation is thought to break down for rough surfaces (\citet{koch}) and at close approach where the continuum approximation of the Navier-Stokes theory fails. Note that Eq.~\ref{eq:SmolDOI} can also be derived through non-Rayleighian methods, such as those applied in \cite{doimain} and in \cite{dhontbriels}.

\subsection{Incorporating Friction}
To incorporate inter-rod friction into the Rayleighian, our ansatz will be to include an additional contribution $\Phi_{\text{fric}}$ to the dissipation  function, yielding
\begin{subequations}
\begin{align} \label{gen-tot-ray}
\mathcal{R} & = \Phi_{\text{drag}} + \Phi_{\text{fric}}(\vec{\omega}) + \dot{A} \\
& = \mathcal{R}_{0} + \Phi_{\text{fric}}(\vec{\omega}).
\end{align}
\end{subequations}

\section{Contact dynamics}
\begin{figure}
\centering\includegraphics[width=0.4\textwidth]{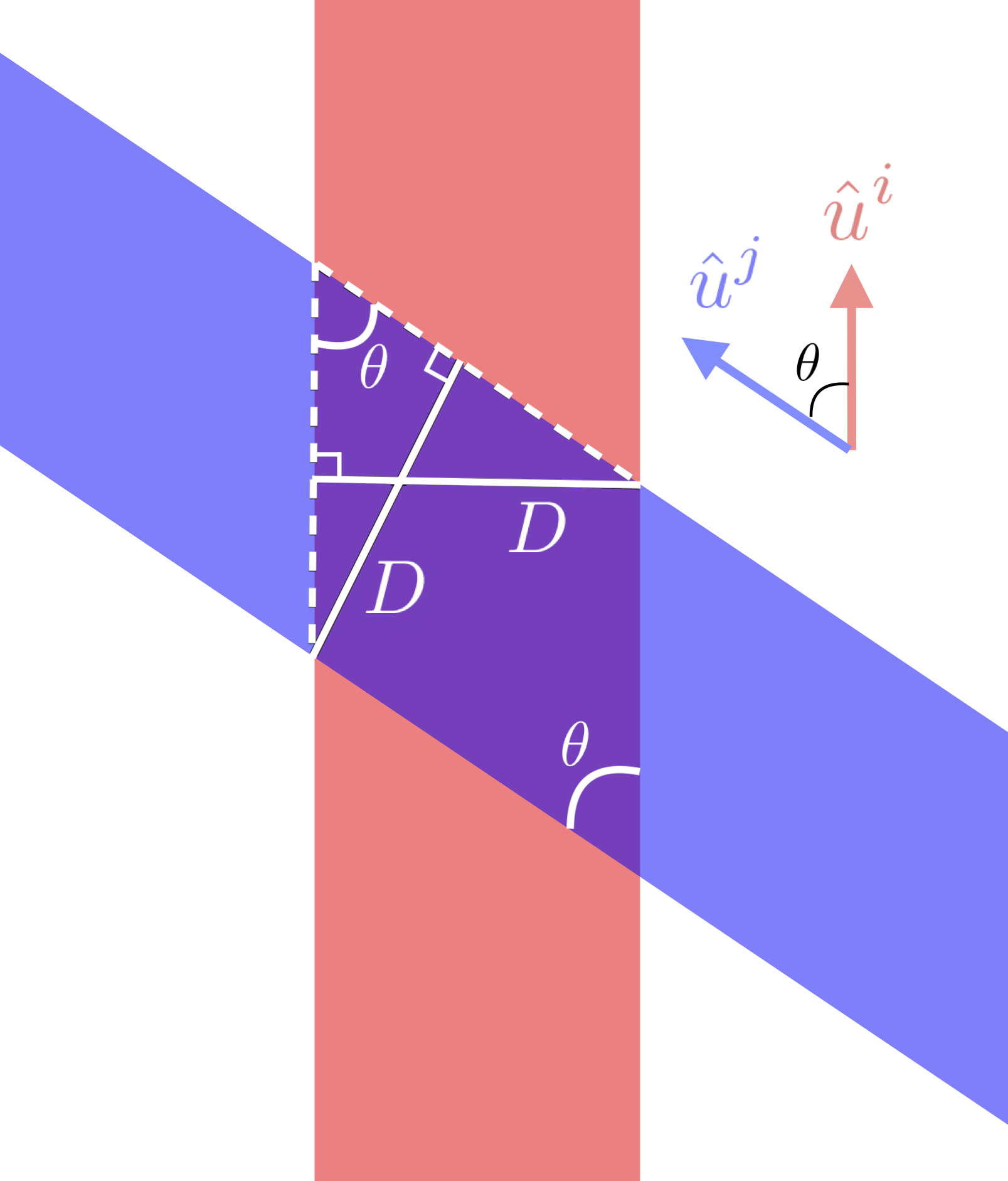}
\caption{Contact area for two rods. Each of the dotted lines has length $D/|\hat{\vec{u}}^{i} \times \hat{\vec{u}}^{j}|$, so the area is $ D^{2}/|\hat{\vec{u}}^{i} \times \hat{\vec{u}}^{j}|$.} \label{fig:rod-area-overlap}
\end{figure}
We assume that all rods' surfaces in contact are in relative motion with a sliding frictional force that dissipates energy. By nature of the Rayleighian method, contacts interacting through static friction cannot be properly accounted for in the resulting dynamical equation. While these contacts do impart torques on rods, they do not contribute directly to the Rayleighian as they do not dissipate energy. 

\subsection{Contact Velocity}
\begin{figure} \centering\includegraphics[width=0.45\textwidth]{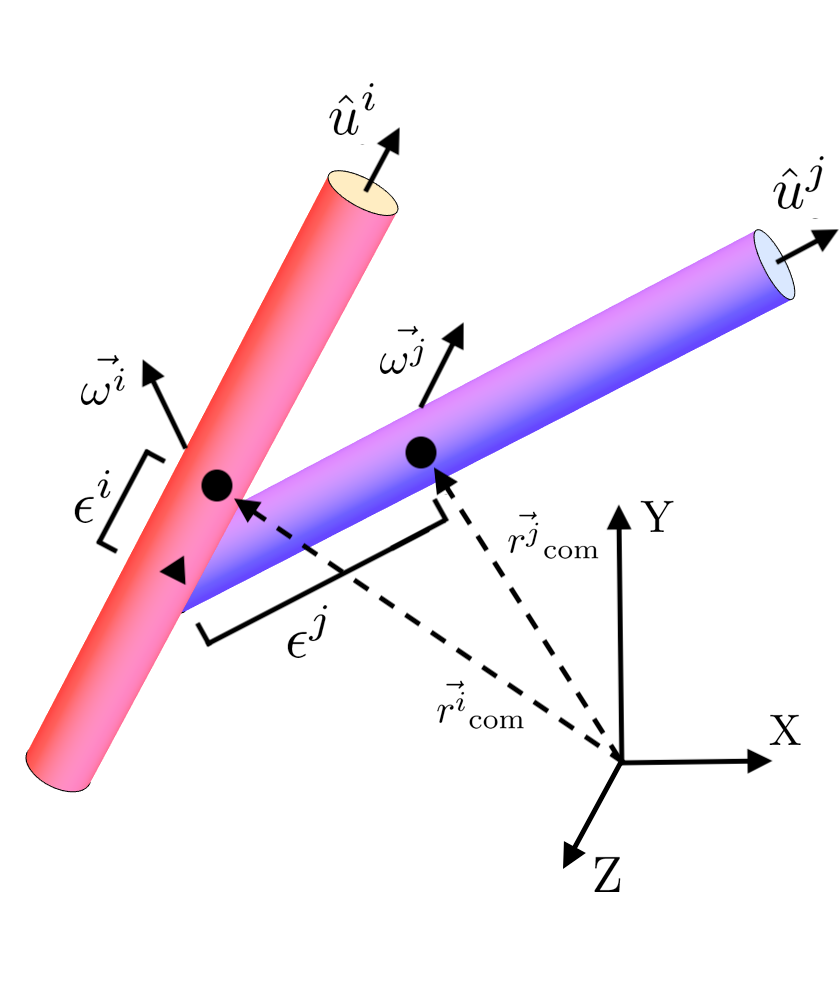}
\caption{Two rods in contact, the black dots are their centers of mass, the black triangle is their contact point. $\epsilon^{i}$ and $\epsilon^{j}$ represent the distance along rod center lines to this point.}\label{fig:rods-contacting}
\end{figure}
Consider two rods $i,j$ in contact, with orientation unit vectors $\hat{\vec{u}}^{i}, \hat{\vec{u}}^{j}$, angular velocity vectors $\vec{\omega}^{i}, \vec{\omega}^{j}$, and positions of center of mass $\vec{r^j}_{\text{com}},\vec{r^i}_{\text{com}}$ as shown in Fig.~\ref{fig:rods-contacting}. These rods are placed in a velocity gradient $\nabla \vec{v}$ and the relative velocity $\vec{v}^{i} - \vec{v}^{j} = \Delta \vec{v}^{ij}$ of the $i$th rod relative to rod $j$ at the point of contact is
\begin{equation} \label{v-c}
\Delta \vec{v}^{ij} =  \vec{\omega^i} \times \epsilon^i \hat{\vec{u}}^i - \vec{\omega^j} \times \epsilon^j \hat{\vec{u}}^j + \nabla \vec{v} \cdot \left(\vec{r^i}_{\text{com}} - \vec{r^j}_{\text{com}} \right).
\end{equation}
Here, $\epsilon$ is the point along the centerline of a rod where the rods touch, which satisfies
\begin{equation} \label{epsilon}
-\frac{L}{2} \leq \epsilon \leq \frac{L}{2}.
\end{equation}
We define a ``contact" by the condition
\begin{equation} \label{contact-condition}
\vec{r^i}_{\text{com}} + \epsilon^i \hat{\vec{u}}^i = \vec{r^j}_{\text{com}} +  \epsilon^j \hat{\vec{u}}^j.
\end{equation}
We have assumed that the rods have high aspect ratio, so the rod diameter is ignored in the contact condition (Eq.~\ref{contact-condition}). This allows us to write the relative velocity $\Delta \vec{v}^{ij}$ as
\begin{equation} \label{full-rel-v}
  \Delta \vec{v}^{ij} =  \vec{\omega^i} \times \epsilon^i \hat{\vec{u}}^i - \vec{\omega^j} \times \epsilon^j \hat{\vec{u}}^j + \nabla \vec{v} \cdot \left(\epsilon^j \hat{\vec{u}}^j - \epsilon^i \hat{\vec{u}}^i \right).
\end{equation}
Friction, and its associated dissipation, is determined by the contact velocity $\vec{v}^{ij}_c$, which is the component of $\Delta \vec{v}^{ij}$ in the plane perpendicular to
\begin{equation} \label{tau}
\vec{\tau}^{ij} = \hat{\vec{u}}^{i} \times \hat{\vec{u}}^{j},
\end{equation}
so
\begin{equation}
  \vec{v}^{ij}_c = \Delta \vec{v}^{ij} \cdot \left( \hat{\tens{I}} - \frac{\vec{\tau}^{ij}\vec{\tau}^{ij}}{|\vec{\tau}|^{2}}\right).
\end{equation}
 \citet{bounoua} proposed that 
\begin{equation} \label{v-c-deltav-equiv}
|\vec{v}^{ij}_c| \approx \left| \Delta \vec{v}^{ij} \right|,
\end{equation}
which is equivalent to assuming very small normal velocities between rods. In dense suspensions of hard rods where there is not much room for inter-rod motion normal to their surfaces this seems a reasonable approximation, which we also make here.
\color{black}

\subsection{Frictional Forms}
\color{black}
We will consider two forms of sliding friction, as done in \cite{bounoua},  and shown by \citet{petrich} to accurately model the interaction of rods sliding past each other in close proximity. First, there is \textit{boundary lubrication} (BL), due to multiple microscopically contacting asperities and trapped fluid, which is proportional to the (tangential) contact velocity $\vec{v}_{c}$. This is distinct from the  gap-dependent  hydrodynamic lubrication (HL) due to shear in a fully-lubricating film. Importantly, we assume that the rods under consideration are not perfect cylinders but are rough, so that contact is indeed between their asperities. This provides a non-zero contact area. BL contact friction has the form
\begin{equation} \label{BL}
\vec{F}^{ij}_{\text{BL}} = - \frac{\eta_{s} D}{|\hat{\vec{u}}^{i} \times \hat{\vec{u}}^{j}|} \vec{v}^{ij}_{c},
\end{equation}
where superscript $i,j$ refers to particles $i$ and $j$, and $\eta_{s}$ is a viscosity proportional to that of the suspending fluid. The factor $|\hat{\vec{u}}^{i} \times \hat{\vec{u}}^{j}|$  in the denominator \footnote{This form only applies for non-perfect order ($S \neq 1$), which we ignore here for simplicity. To account for this one can use instead $\vec{F}^{ij}_{\text{BL}} = - \eta_{s} D\,\, \textrm{min}\left(\frac{L}{D},\frac{1}{|\hat{\vec{u}}^i\times\hat{\vec{u}}^j|}\right)\vec{v}^{ij}_{c}$.} of Eq.~\ref{BL} accounts for the scaling of friction with contact area, as in \citet{yamane} and \cite{djalili}. As shown in Fig.~\ref{fig:rod-area-overlap}, the contact area between  two rods $i$ and $j$ is $D^{2}/|\hat{\vec{u}}^{i} \times \hat{\vec{u}}^{j}|$. \footnote{In the numerator of Eq.~\ref{BL} we have a power of the area of overlap and of the relevant rate, the shear rate, which is proportional to $\vec{v}_{c}/D$, so we are left with only one power of $D$.}

The other form of friction is solid Amontons-Coulomb friction,
\begin{equation} \label{sol}
\vec{F}^{ij}_{\text{solid}} = -\mu_{k} |\vec{F}_{\text{N}}| \hat{\vec{v}}^{ij}_{c},
\end{equation}
which only depends on the direction of the contact velocity, and not its magnitude. Here, $\mu_{k}$ the coefficient of kinetic friction, and $|\vec{F}_{\text{N}}|$ the normal force between the sliding surfaces. We assume that surfaces in relative motion are subject to a constant coefficient of friction.

By integrating the power associated with lubricated friction we find a dissipation function with the familiar quadratic form
\begin{equation} \label{lub-dissi}
\Phi_{\text{BL}}^{ij} = \frac{\eta_{s} D}{2|\hat{\vec{u}}^{i} \times \hat{\vec{u}}^{j}|} \left(v_{c}^{ij}\right)^2,
\end{equation}
which corresponds to the dissipation due to a single pair of rods in contact (see \citet{yamane} for an example of a pairwise dissipation function accounting for hydrodynamic lubrication).
Similarly, we can integrate the power associated with the frictional force to obtain a different form for the dissipation function (\citet{sonnet}, \cite{virga2014comment}):
\begin{equation} \label{solid-dissi}
\Phi_{\text{sol}}^{ij} = \mu_{k}|\vec{F}_{\text{N}}| |\vec{v}_{c}^{ij}|.
\end{equation}
This construction maintains a positive semi-definite dissipation function (a consequence of Onsager's principle \cite{doionsager}), as required. We will average the  pair-wise frictional dissipation  over all contacts for a solution of volume fraction $\phi$ in §~\ref{sec:contact-field}; and calculate the Smoluchowski Equation for the distribution function $\psi(\uu)$ of rod orientations in §~\ref{sec:Smol}.

\subsection{Contact Torque}
The torque on the fluid from the $i$th rod associated with a solid frictional contact between rods $i$ and $j$ is given, from Eqs.~(\ref{eq:torque}, \ref{solid-dissi}, \ref{full-rel-v}), by
\begin{subequations}
\label{solid-ray-variate}
\begin{align} 
 \vec{\Gamma}^{ij}_{\text{solid}}
& =  -\mu_{k}|\vec{F}_{\text{N}}| \hat{\vec{v}}_{c}^{ij} \cdot  \frac{\partial}{\partial \vec{\omega}^{i}} \vec{v}^{ij}_c \\
& = -\mu_{k}|\vec{F}_{\text{N}}| \hat{\vec{v}}_{c}^{ij} \times \left(\epsilon^{i} \hat{\vec{u}}^i \right).
\end{align}
\end{subequations}
Using Eq.~\ref{lub-dissi}, for a lubricated contact we have simply
\begin{align} \label{lub-ray-variate}
 \vec{\Gamma}^{ij}_{\text{BL}}
& =  -\frac{\eta_{s} D}{|\hat{\vec{u}}^{i} \times \hat{\vec{u}}^{j}|} \vec{v}_{c}^{ij} \times \left(\epsilon^{i} \hat{\vec{u}}^i \right).
\end{align}
\color{black}

\section{Contacts as a Mean Field} \label{sec:contact-field}
A given rod in a dense suspension has some number of ``contact neighbors" $c$. The probability that some rod contacts another, ${\cal P}_{\textrm{con}}$, can be used to calculate the average of $c$,
\begin{equation}  \label{con-prob}
\mean{c} = \int \int \psi^jd\uu^j \psi^id\uu^i \,\,{\cal P}_{\textrm{con}}\left(\hat{\vec{u}}^i,\hat{\vec{u}}^j \right).
\end{equation}
In the above, we have specified that ${\cal P}_{\textrm{con}}$ depend only on rod orientation, but in principle it also depends on $\vec{r^i}_{\text{com}}$ and $\vec{r^j}_{\text{com}}$. However, here we consider a spatially homogeneous suspension, so we need not average over spatial coordinates.
\citet{doi1978}, in calculating the number of tube intersections in a tube model for polymers, first pointed out that for large aspect ratio rods ${\cal P}_{\textrm{con}}(\uu^i,\uu^j)$ is proportional to $\vec{\tau}^{ij}=|\uu^i\times\uu^j|$ (Eq.~\ref{tau}). The excluded volume between two rods of angle $\sin\theta_{ij}=\vec{\tau}^{ij}=|\uu^i\times\uu^j|$ was given by \citet{onsager1949}, 
\begin{equation}
    v_{ex}^{ij} = L^2D \left|\vec{\tau}^{ij} \right|.
\end{equation}
Hence, we can estimate the contact probability as the ratio of excluded volume to the average volume per rod, 
\begin{equation}
{\cal P}_{\textrm{con}}\left(\hat{\vec{u}}^i,\hat{\vec{u}}^j\right) = \bar{\rho} v_{ex}^{ij}
\end{equation}
where $\bar{\rho}=4\phi/(\pi D^2L)$ is the number of rods per volume.

By using this form for ${\cal P_{\textrm{con}}}$, 
Philipse \cite{randomcontact} derived the ``random contact equation" for the isotropic case $S=0$, where $\psi^i=1/(4\pi)$: 
\begin{equation} \label{rand_contact_eq}
\begin{split}
\langle c \rangle & \approx \frac{\bar{\rho}}{2} \langle v_{\text{ex}} \rangle_{S=0}\\
& = \phi\frac{L}{D}.
\end{split}
\end{equation}
For $S \neq 0$ we find
\begin{subequations} \label{lc-contacts-pre}
\begin{align}
\langle c \rangle & \approx \frac{\bar{\rho}}{2} \langle v_{\text{ex}} \rangle \\
& = \bar{\rho} DL^{2} \left\langle |\vec{\tau}| \right\rangle \\
& = \phi \frac{L}{D} \frac{4}{\pi} \int{ \int{ |\vec{\tau}| \psi^{i} \text{d}\hat{\vec{u}}^i } \psi^{j} \text{d}\hat{\vec{u}}^j } \label{lc-contacts0} 
\end{align}
\end{subequations}
Eq.~\ref{lc-contacts0} is equivalent to that derived by Toll \cite{toll}, save for his inclusion of a contribution from end-end contacts.
The average over $|\vec{\tau}|=|\uu^i\times\uu^j|$ is not possible in closed form and depends on the orientational distribution function $\psi$. Following Doi's approximation (see Eq.~\ref{doi-tau}), we obtain
\begin{equation}\label{lc-contacts}
\mean{c} \approx \phi \frac{L}{D}\left(1-\tfrac{2}{3}S^2\right),
\end{equation}
where all dependence on higher moments of $\tens{Q}$ have been ignored  and a uniaxial form assumed.  Perfectly ordered rods should  only have contacts with neighbors in the plane perpendicular to the director $\uvec{n} $~\footnote{Including the $S=1$ contribution would give the modified form $\mean{c} = \phi \frac{L}{D}\left(1-S^2\right) + C_a(\phi,L)\delta(S-1)$, where $C_a$ is equal to the number of contacts between discs in the plane.}, so that in the limit S=1 the number of contacts should not scale with the aspect ratio $L/D$. Hence, we approximate the number of contacts per rod in  an ordered system as
\begin{equation} \label{final-c-s}
\mean{c} = \phi \frac{L}{D}\left(1-S^2\right).
\end{equation}
between discs in a plane. 
\color{black}

Given that the frictional dissipation for some rod $i$ scales with the number of contacts for that rod, we can write it as the sum over these contacts and form an average similar to Eq.~\ref{lc-contacts-pre};
\begin{subequations}
\begin{align}
\Phi^{i} &= \sum_{j\in c} \Phi^{ij} \label{basic-tot-ray} \\ \label{broad-avg}
&= \int \psi^jd\uu^j\,\, {\Phi}^{ij} \,\,{\cal P}_{\textrm{con}} \left(\hat{\vec{u}}^i,\hat{\vec{u}}^j \right).
\end{align}
\end{subequations}
To form a mean field dissipation function $\Phi^{\text{mf},i}$, we must also incorporate the torque arms $\epsilon_{i}$ and $\epsilon_{j}$ into our average (see Eqs.~\ref{full-rel-v}, ~\ref{v-c-deltav-equiv} and ~\ref{solid-dissi}). We assume their distribution is uniform along the length (Eq.~\ref{epsilon}) and independent of rod orientation. So,
\begin{subequations}
\begin{align}\label{mean-field}
\Phi^{\text{mf},i}&\left(\nabla \vec{v}, \hat{\vec{u}}^{i},\vec{\omega}^{i} \right)  = \left\langle \Phi^{ij} \right\rangle_{j} \\[6.0truept]
 &\equiv \phi \frac{L}{D}\frac{4}{\pi}\int\displaylimits_{-L/2}^{L/2}\frac{\text{d}\epsilon^{i}}{L}\int\displaylimits_{-L/2}^{L/2}\frac{\text{d}\epsilon^{j}}{L} \int{|\vec{\tau}| \psi^{j} \text{d}\hat{\vec{u}}^{j}} \Phi^{ij},
\end{align}
\end{subequations}
where the power of $\frac{1}{L^2}$ normalizes the $\epsilon$ integrals. This method of averaging quantities in a suspension of rods was used in \cite{djalili,bounoua,lubefric}. From Eq.~\ref{broad-avg} and Eq.~\ref{con-prob} we see that, if we assume a uniform dissipation per contact $\bar{\Phi}^{i}$, this expression can be approximated as
\begin{equation} \label{mean-mean-field}
\Phi^{\text{mf},i}\approx
 {\mean{c}}\bar{\Phi}^{i},
\end{equation}
\emph{i.e.} 
the mean number of contacts each with uniform dissipation $\bar{\Phi}^{i}$.
\section{Dissipation Functions and Smoluchowski Equations}\label{sec:Smol}

\subsection{Contribution from Solid Contacts}

Following Eq.~\ref{mean-field}, the mean field form of the dissipation function for solid frictional contacts can be obtained by inserting Eq.~\ref{solid-dissi} and also averaging over $\hat{\vec{u}}^i$,
\begin{subequations} \label{full-avg-solid}
\begin{align}
\Phi_{\text{sol}}^{\text{mf}} & =  \int \psi^{i} \text{d}\hat{\vec{u}}^{i} \left\langle \mu_{k}|\vec{F}_{\text{N}}|  |\vec{v}_{c}| \right\rangle_{j} \\
& \equiv \mu_{k}|\vec{F}_{\text{N}}|  \langle \langle |\vec{v}_{c}| \rangle \rangle,
\end{align}
\end{subequations}
where the inner set of brackets is defined by Eq.~\ref{mean-field}, and the second pair of angle brackets denotes the additional $\hat{\vec{u}}^{i}$ integral. The second line in Eq.~\ref{full-avg-solid} follows from the assumption that $\mu_{k}$ and $|\vec{F}_{\text{N}}|$ are independent of $\hat{\vec{u}}^{i},\hat{\vec{u}}^{j},  \epsilon^{i},$ and $\epsilon^{j}$.
The torque associated with Eq.~\ref{full-avg-solid} follows from Eq.~\ref{solid-ray-variate}, now averaged over the variables associated with rod $i$'s contact neighbors:
\begin{subequations}
\begin{align}
\vec{\Gamma}_{\text{sol}}^{\text{mf},i} & = \frac{\delta }{\delta \vec{\omega}_{i}} \Phi_{\text{sol}}^{\text{mf}} \\
& = -\mu_{k} |F_{\text{N}}| \left \langle  \hat{\vec{v}}_{c}^{ij} \times \epsilon^{i} \hat{\vec{u}}^i  \right\rangle_{j} \\
 & = -\mu_{k} |F_{\text{N}}| \left \langle  \frac{\left(\epsilon^{i} \right)^2}{|\vec{v}_{c}|} \left[ (\vec{\omega}^{i} \times \hat{\vec{u}}^{i}) \times \uu^{i} + \vec{\omega}_{0} \right] \right\rangle_{j} \\
 & = \mu_{k} |F_{\text{N}}| \left[ \vec{\omega}^{i} - \vec{\omega}_{0} \right] \left\langle  \frac{\left(\epsilon^{i} \right)^2}{|\vec{v}_{c}|} \right\rangle_{j}, \label{torque-mf-fric}
\end{align}
\end{subequations}
where we have used Eqs.~\ref{full-rel-v} and ~\ref{flow-omega} in the third line, the brackets are defined by Eq.~\ref{mean-field}, and the fourth line follows from $(\vec{\omega}^{i} \times \hat{\vec{u}}^{i}) \times \uu^{i} = -\vec{\omega}^{i}$. Terms odd in $\epsilon_{i}$ and $\epsilon_{j}$ have been dropped due to the assumed symmetry in their distribution. Note that the contact velocity $\vec{v}_c$ depends on both $\vec{\omega}_i$ and $\vec{\omega}_j$, as in Eq.~\ref{full-rel-v}.

The form of Eq.~\ref{torque-mf-fric} differs from the results of \citet{bounoua} and \citet{sandstrom}, who calculated a vanishing average torque due to contacts. They assumed that the equation of motion is given by the sum of the dynamics in the dilute case, Jeffery's equation, to which they simply added a torque due to frictional contacts in the semi-dilute regime. Both \cite{bounoua,sandstrom} obtained an average frictional torque that depended on odd powers of $\epsilon^{i}$ and $\epsilon^{j}$ (specifically in the large aspect ratio limit, see Appendix \ref{appendix:fiber-torque}) inside the same average as Eq.~\ref{mean-field}, which vanishes when integrating over the contact position $\epsilon^i$. By contrast, we did not assume that the equation of motion (analogous to $\vec{\omega}_{\textrm{min}}$) is a Jeffery-like equation with an additional torque. Rather, we \textit{derived} the dynamics based on a dissipation function that incorporates (solid or lubricated) frictional drag. This yields an $\vec{\omega}_{\text{min}}$ for the rods, with an associated torque, after taking the variation. Our approach necessarily results in even powers of $\epsilon^i$ (Eq.~\ref{torque-mf-fric}), whose average over all contacts is non-zero. Hence, as $\left \langle  \frac{(\epsilon^{i})^2}{|\vec{v}_{c}|} \right\rangle_{j} \neq 0$, the solid frictional torque is non-zero and so is its contribution in the Smoluchowski equation.
\subsection{Contribution from Boundary Lubricated Contacts}
For lubricated contacts (Eq.~\ref{BL}), the averaged form of the corresponding dissipation function is
\begin{equation} \label{full-avg-lub}
\Phi_{\text{BL}}^{\text{mf}} = \frac{\eta_{s} D}{2}  \left\langle \left\langle \frac{{v}_{c}^2}{|\vec{\tau}|} \right\rangle \right\rangle,
\end{equation}
where the double brackets are as in Eq.~\ref{full-avg-solid}.
Employing Eq.~\ref{lub-ray-variate}, the associated torque on the fluid is
\begin{widetext}
\begin{subequations}
\begin{align}
\vec{\Gamma}^{\text{mf,i}}_{\text{BL}} & = \frac{\delta }{\delta \vec{\omega}_{i}} \Phi_{\text{BL}}^{\text{mf}} \\
& = -\frac{4\eta_{s}\phi}{\pi L}  \int_{-\frac{L}{2}}^{\frac{L}{2}}{\text{d}\epsilon^{i}}\int_{-\frac{L}{2}}^{\frac{L}{2}}\text{d}\epsilon^{j} \int{ \psi^{j}\text{d}\hat{\vec{u}}^{j}} (\vec{v}_{c}^{ij} \times \epsilon^{i} \hat{\vec{u}}^{i}) \\
& = \frac{4\eta_{s}\phi}{\pi L} \int_{-\frac{L}{2}}^{\frac{L}{2}}{\text{d}\epsilon^{i}}\int_{-\frac{L}{2}}^{\frac{L}{2}}\text{d}\epsilon^{j} \int{ \psi^{j}\text{d}\hat{\vec{u}}^{j}}  \left(\epsilon^{i} \right)^2 \left[ \vec{\omega}^{i} -  \vec{\omega}_{0} \right] \\
& = \frac{\eta_{s}\phi L^3}{ 3 \pi} \left[ \vec{\omega}^{i} - \vec{\omega}_{0} \right] \label{torque-mf-lub}
\end{align}
\end{subequations}
\end{widetext}
where again terms odd in $\epsilon_{i}$ and $\epsilon_{j}$ have been dropped, and in the second line the weighting of $|\vec{\tau}|$ of the contact average canceled out with a power in the denominator introduced in Eq.~\ref{BL}.
\subsection{Smoluchowski Equation} \label{subsec:new-smol}
The Rayleighian (per rod) for a suspension under shear\textit{ without} considering  frictional contacts (Eqs.~\ref{fe-dissi}, \ref{shear-dissi}, \ref{rayl-def}) is 
\begin{equation} \label{full-flow-rayl}
\begin{split}
\mathcal{R}_{0} = \int\psi^i  \text{d}\hat{\vec{u}}^i & \left[ \frac{\zeta_{r}}{2} (\vec{\omega}^i - \vec{\omega}_{0})^2 + \frac{\zeta_{r}}{4} (\uu^{i} \cdot \nabla \vec{v} \cdot \uu^{i})^2  \right.\\
&\quad\left. + \vec{\omega}^i \cdot \Rhat\tilde{U} \right]  + \frac{\eta_{s}}{\bar{\rho}} \tens{D}:\tens{D} ,
\end{split}
\end{equation}
where $\tilde{U} = \kbT \ln{\psi} + U\left(\hat{\vec{u}}^i \right)$ is the combination of Brownian and mean field excluded volume potentials, $\tens{D} = 1/2\left(\nabla \vec{v} + \nabla \vec{v}^{\text{T}}\right)$ is the symmetric velocity gradient tensor, $\bar{\rho}$ is the mean particle density, and $\vec{\omega}_{0}$ is defined by Eq.~\ref{flow-omega}. The total Rayleighian to minimize is given by Eq.~\ref{gen-tot-ray}, which is the sum of ${\cal R}_0$ and the dissipation from lubricated and solid frictional contacts:
\begin{equation}
    \mathcal{R} = \mathcal{R}_{0} + \Phi_{\text{sol}}^{\text{mf}} + \Phi_{\text{BL}}^{\text{mf}}.
\end{equation}
Finally, we follow Wyart and Cates \cite{wyartcates} and presume that some fraction $\Theta$ of the $\langle c \rangle$ contacts are solid and the remaining fraction $1-\Theta$ are lubricated. The function $\Theta$ depends on the ratio of the typical normal force $|F_N|$ to a characteristic force $F^{\ast}$ required to break the lubrication film\footnote{Note the implication of Eq.~\ref{press-def} that $\vec{F}_{N} = \vec{F}_{N}(\boldsymbol{\sigma})$, introducing a self-consistency in determining $\boldsymbol{\sigma}$, given Eq.~\ref{sig_main_fric}}. We then assume that the average normal force is proportional to the fluid pressure $P$:
\begin{subequations} \label{press-def}
\begin{align}
    \frac{|\vec{F}_{N}|}{A_{c}} & \simeq P \\
    & = -\frac{1}{3}\text{Tr}(\boldsymbol{\sigma}) ,
\end{align}
\end{subequations}
where $A_{c}$ is the cross-sectional area per contact. Hence, we assume $\Theta$ is a function of the reduced pressure $p\equiv P/P^{\ast}$,
where the characteristic pressure $P^{*}$ is given by
\begin{equation}
    P^{*} = \frac{F^{*}}{A_{c}}. 
\end{equation}
The area per contact scales with the number of contacts per volume $N_{c}$, and from $N_{c} = 4 \phi \langle c \rangle/(\pi D^{2} L)$ we have
\begin{subequations} \label{ac-def}
\begin{align}
    A_{c} & \sim N_{c}^{-\frac{2}{3}} \\
    & = \left(\frac{4 (1- S^2)}{\pi} \right)^{-\frac{2}{3}} \frac{D^2}{\phi^{\frac{4}{3}}},
\end{align}
\end{subequations}
where we have inserted Eq.~\ref{final-c-s} in the second line. Candidates for $F^{*}$ include electric double layer forces, which have been used in simulations of frictional contacts by \citet{setomorris}, steric repulsion from polymeric brushes grafted on to the surface of particles \cite{IsaSpencerStribeck_PRL2013}, and plastic deformation of surface asperities \cite{FieldingPlasticFriction_PRL2023}. We assume that $\Theta$ has the following limiting behavior:
\begin{equation} \label{contact-split}
\Theta(p) =
\bigg\{
    \begin{array}{lr}
        1, & \text{for } p \gg 1\\
        0, & \text{if } p \ll 1,
    \end{array}
\end{equation}
and varies sharply from 0 to 1 at $P=P^{*}$. Hence, \begin{equation} \label{tot-rayl-both}
    \mathcal{R} = \mathcal{R}_{0} + \left[1-\Theta\left(p\right)\right]\Phi_{\text{BL}}^{\text{mf}} + \Theta(p)\Phi_{\text{sol}}^{\text{mf}}.
\end{equation} 
Upon minimization with respect to $\vec{\omega}^{i}$ (see Eqs.~\ref{torque-mf-fric}, \ref{torque-mf-lub}), we find
\begin{equation} \label{full-omega-min}
\vec{\omega}^{i}_{\text{min}} =  \vec{\omega}_{0} - \frac{\Rhat\tilde{U}}{\zeta_{r} + \zeta_F(\vec{F}_{N},\vec{v}_{c})},
\end{equation}
where 
\begin{equation} \label{eq:zeta_f}
\zeta_F (F_N,\vec{v}_{c})= (1-\Theta(p))\frac{\eta_{s}\phi L^3}{ 3 \pi} + \Theta(p) \mu_{k} |\vec{F}_{N}| \left \langle  \frac{(\epsilon^{i})^2}{|\vec{v}_{c}(\vec{\omega}_{\textrm{min}})|} \right\rangle_{j} 
\end{equation}
is the rotational drag function due to friction, which depends on the stress through $F_N$ and the local rotational dynamics through $|\vec{v}_c|$. Eq.~\ref{full-omega-min} is a self-consistent equation of motion for  $\vec{\omega}_{\textrm{min}}$, since the drag force itself depends on the rotation rate $\vec{\omega}_{\textrm{min}}$.

The corresponding Smoluchowski equation is 
\begin{equation}\label{eq:SmolNew}
    \dot{\psi}= 
    -\Rhat \cdot \left( \vec{\omega}_{0} \psi \right) +
    \Rhat\cdot\left[\frac{\psi\Rhat\tilde{U}}{\zeta_{r} +\zeta_F (\vec{F}_{N},\vec{v}_{c}))}\right].
\end{equation}
This is a mean field equation, analogous to the Doi model near the isotropic-nematic transition \cite{doiedwards}, which depends on the average value of the order parameter in the excluded volume potential $\tilde{U}$), as $\zeta_{F}$ (Eq.~\ref{eq:zeta_f}) is defined in terms of an average over $\psi$. To resolve Eq.~\ref{eq:SmolNew}'s dependence on this average in a numerical calculation, one could apply the technique of Larson \cite{larson_arrested} and expand $\psi$ in spherical harmonics.

\section{Predictions from Smoluchowski Equation} \label{sec:predictions}

\subsection{Lubricated Contact Limit}
For only lubricated contacts ($\Theta(p) = 0$), Eq.~\ref{eq:zeta_f} simplifies considerably and $\vec{\omega}_{\text{min}}$ is exactly the same form as Eq.~\ref{doi-rayl-flow-omega},
\begin{equation} \label{lub-omega}
\vec{\omega}_{\text{min}}=  \vec{\omega}_{0} -\frac{1}{\tilde{\zeta}_{r}}\Rhat \tilde{U},
\end{equation}
but now with the renormalized rotational drag constant
\begin{equation} \label{eq:lub_contacts_drag}
    \tilde{\zeta}_{r} = \zeta_{r} + \frac{\eta_{s}\phi L^3}{ 3 \pi}.
\end{equation}
As this quantity is independent of $\hat{\vec{u}}^{i}$, the resulting Smoluchowski equation has the same form as that of Doi-Edwards (See Eq.~\ref{eq:SmolDOI}), as does $\dot{\tens{Q}}$. The effect of lubricated contacts is to increase $\zeta_{r}$; \emph{i.e.}, the rate of rotation of rods is decreased by the presence of contacts, as expected.  

\subsection{Solid Contact Limit}
For all solid-like contacts ($\Theta(p) = 1$), we obtain
\begin{equation} \label{pre-scale_solid_omega}
\vec{\omega}^{i}_{\text{min}} =  \vec{\omega}_{0} - \frac{\Rhat\tilde{U}}{\zeta_{r} + \mu_{k} |\vec{F}_{N}| \left \langle  \frac{(\epsilon^{i})^2}{|\vec{v}_{c}(\vec{\omega}_{\textrm{min}})|} \right\rangle_{j} },
\end{equation}
where now the denominator on the RHS is a function of $\vec{\omega}^{i}$, so that the equation is self-consistent (The angular velocity dependence of $\vec{v}_c$ is shown in Eq.~\ref{full-rel-v}$)$. This is not an unexpected result; the dependence of  solid friction on the \textbf{direction} of $\vec{v}_{c}$, but not its magnitude, leads to an associated torque (Eq.~\ref{torque-mf-fric}) that is a non-linear function of $\vec{\omega}$. Similarly, the Smoluchowski equation (Eq.~\ref{eq:SmolNew}) will be self-consistent, as the denominator on the RHS depends on $\psi$.

We choose to scale the rod length $L$ out of $\epsilon^{i}$, and a  characteristic relative speed $L \dot{\gamma}$ between rod contacts out of the contact velocity (since the dominant motion leading to frictional torque is possibly inter-rod rotation), defining
\begin{subequations}
\begin{align}
    \hat{\epsilon}^{i} & = \frac{1}{L}  \epsilon^{i}, \\
    |\hat{\vec{v}}_{c}| & = \frac{1}{L \dot{\gamma}} |\vec{v}_{c}|.
\end{align}
\end{subequations}
Then we can write Eq.~\ref{pre-scale_solid_omega} as
\begin{equation} \label{pre-taylor-omega-min}
    \vec{\omega}^{i}_{\text{min}} =  \vec{\omega}_{0} - \frac{\Rhat\tilde{U}}{\zeta_{r} \left[1+ \Delta \left\langle \frac{(\hat{\epsilon}^{i})^2}{|\hat{\vec{v}}_{c}|} \right\rangle_{j} \right]}, 
\end{equation}
where the ratio
\begin{equation} \label{delta}
\Delta = \frac{\mu_{k} |\vec{F}_{N}| L}{\zeta_{r} \dot{\gamma}}.
\end{equation}
compares the characteristic torque from solid friction to that of hydrodynamic drag.
Hence, solid frictional contacts increases the effective rotational drag constant and brings $\vec{\omega}_{\text{min}}$ closer to Jeffery's equation $\vec{\omega}^{i}_{\text{min}} =  \vec{\omega}_{0}$. 

The ratio $\Delta$ does not diverge as $\dot{\gamma}\rightarrow0$, as the normal force $F_N$ between rods is induced by flow. This force $F_N$ should contain a component of the shear stress, so we expect that $F_N\simeq \eta A_{c} \dot{\gamma}(1 + {\cal O}(\dot{\gamma}) + \ldots$), where $\eta$ is the viscosity and $A_{c}$ is the characteristic area between frictional contacts. The rotational drag coefficient scales as $\zeta_r\sim \eta_sL^3$ \cite{doiedwards}, up to a numerical prefactor and a weak logarithmic dependence on $L/D$, so along with $|F_{N}(\dot{\gamma})|$,
\begin{equation}
    \Delta\propto\frac{\mu \eta A_c}{\eta_s L^2}.
\end{equation}
Using Eq.~\ref{ac-def}, we have 
\begin{equation}
    \Delta\propto\mu\frac{\eta}{\eta_s}\left(\frac{D}{L \phi^{\frac{2}{3}}}\right)^2\left(\frac{1}{1-S^2}\right)^{2/3},
\end{equation}
where we have suppressed numerical prefactors.
In the case where the frictional torque is much smaller than the rotational drag torque, $\Delta\ll1$,  we can expand in powers of $\Delta$ to obtain the following  Smoluchowski equation
\begin{align} \label{new-smol}
\dot{\psi} 
    = & -\Rhat \cdot \left\{\psi\vec{\omega}_{0} - \frac{\psi\Rhat\tilde{U}}{\zeta_{r}} \left(1-\Delta \left\langle \frac{(\hat{\epsilon}^{i})^2}{|\hat{\vec{v}}_{c}|} \right\rangle_{j}\right)\right\}.
\end{align}
The first two terms above follow  Doi \cite{doibook}, and the final term is a friction-driven addition, which scales with the magnitude of the average torque due to frictional contacts (Eq.~\ref{torque-mf-fric}), as expected.

\subsubsection{Dynamical equation for $\tens{Q}$} \label{subsec:new-q-tens}
We can use the Smoluchowksi equation as approximated in Eq.~\ref{new-smol} to  calculate the dynamical equation for
the order tensor $\tens{Q}$. Beginning from
\begin{equation}
    \dot{\tens{Q}} = \int \left( \hat{\vec{u}}\hat{\vec{u}} - \frac{\hat{I}}{3} \right) \dot{\psi} \text{d}\hat{\vec{u}}, 
\end{equation}
we find
\begin{align} \label{gen-Q-dot-with-fric}
    \dot{\tens{Q}} & = \dot{\tens{Q}}_{\text{DE}} + \dot{\tens{Q}}_{\text{FRIC}} \\
    & = \dot{\tens{Q}}_{\text{DE}} - \int \left( \hat{\vec{u}}\hat{\vec{u}} - \frac{\hat{I}}{3} \right) \Rhat \cdot \left( \frac{\Rhat\tilde{U}}{\zeta_{r}} \Delta \left\langle \frac{(\hat{\epsilon}^{i})^2}{|\hat{\vec{v}}_{c}|} \right\rangle_{j}   \psi \right) \text{d}\hat{\vec{u}},
\end{align}

where $\dot{\tens{Q}}_{\text{DE}}$ is Doi's \cite{doimain} equation of motion for $\tens{Q}$ and $\dot{\tens{Q}}_{\textrm{FRIC}}$ is due to the effects of friction. This term can be evaluated using the following identity \cite{weinanproof},
\begin{equation}
    \int \left( \hat{\vec{u}}\hat{\vec{u}} - \frac{\hat{I}}{3} \right) \Rhat \cdot \left( \vec{x} \psi \right) \text{d}\hat{\vec{u}} = \int (\hat{\vec{u}} \times \vec{x}) \otimes \hat{\vec{u}} + \hat{\vec{u}} \otimes (\vec{x} \times \hat{\vec{u}}) \psi \text{d}\hat{\vec{u}},
\end{equation}

\section{Frictional contribution to the stress} \label{subsec:new-stress}

We have thus-far constructed the Rayleighian on a per-particle basis. To calculate the stress tensor $\boldsymbol{\sigma}$ we need the Rayleighian $\mathcal{R}_{v}$ expressed  per volume instead of per rod \cite{doionsager}, 
\begin{equation} \label{rayl-per-vol}
    \mathcal{R}_{v} = \bar{\rho}\,\mathcal{R},
\end{equation}
where we assume a uniform particle density $\bar{\rho}$.
The stress tensor is given by \cite{doibook} \footnote{As we have assumed spatial homogeneity, we need not calculate a variational derivative to find the stress. For a general $\mathcal{R} = \int \rho(r) \mathcal{R}_{\rho} \text{d}r^{3}$, where $\mathcal{R}_{\rho}$ is a per particle Rayleighian that depends on spatial coordinate $r$ and $\rho(r)$ the density at $r$, then $\boldsymbol{\sigma}_{\alpha \beta} = \frac{\delta \mathcal{R}}{\delta (\nabla v)_{\alpha \beta}}$.}
\begin{equation} \label{stress-from-rayl}
\boldsymbol{\sigma}_{\alpha \beta} = \frac{\partial \mathcal{R}_{v}}{\partial (\nabla v)_{\alpha \beta}},
\end{equation}
 and using the Rayleighian ${\cal R\/}$ of Eq.~\ref{tot-rayl-both} we obtain
\begin{equation}\label{eq:stress-total}
\begin{split}
    \boldsymbol{\sigma}_{\alpha \beta} &
    = \bar{\rho}\left[\frac{\partial \mathcal{R}_{0}}{\partial (\nabla v)_{\alpha \beta}} + \Theta(p)\frac{\partial \Phi_{\text{sol}}^{\text{mf}}}{\partial (\nabla v)_{\alpha \beta}}\right. \\[6truept]
    &\qquad\qquad \left.+ (1 - \Theta(p))\frac{\partial \Phi_{\text{BL}}^{\text{mf}}}{\partial (\nabla v)_{\alpha \beta}}\right].
\end{split}
\end{equation}
We can identify $\bar{\rho}\frac{\partial \mathcal{R}_{0}}{\partial (\nabla v)_{\alpha \beta}}$ as the stress due to non-frictional dissipation, up to the insertion of a $\vec{\omega}_{\textrm{min}}$, and the remaining terms in Eq.\ref{eq:stress-total} as that due to frictional dissipation. We show in Appendix~\ref{appendix:doi-stress} that this stress tensor can also be represented as
\begin{widetext}
\begin{equation} \label{sig_combo}
\boldsymbol{\sigma}_{\alpha \beta} = \boldsymbol{\sigma}_{\alpha \beta}^{\textrm{DE}}+
\boldsymbol{\sigma}_{\alpha \beta}^{\textrm{F}},
\end{equation}
where $\boldsymbol{\sigma}_{\alpha \beta}^{\textrm{DE}}$ is the Doi-Edwards stress tensor for rod-like suspensions without friction, and $\boldsymbol{\sigma}_{\alpha \beta}^{\textrm{F}}$ is the additional stress due to frictional contacts:
\begin{equation} \label{sig_main_fric}
\begin{split}
    \boldsymbol{\sigma}_{\alpha \beta}^{\textrm{F}} &= \bar{\rho}\left\langle \left\langle \left[\Theta(p) 2\mu_k|\vec{F}_{N}|\frac{(\epsilon^{i})^2}{|\vec{v}_{c}|}  + \left(1-\Theta\left(p\right)\right) 4\eta_s D \frac{(\epsilon^{i})^2}{|\vec{\tau}|} \right] \left[\uu^{i}_{\alpha}\uu^{i}_{\beta}\,\left(\uu^{i}\!\cdot\!\vec{\nabla} \vec{v}\!\cdot\!\uu^{i}\right)
    +\vec{f} \cdot \frac{\partial \vec{f}}{\partial (\nabla v)_{\alpha \beta}}\right]  \right\rangle
    \right\rangle \\[5truept]
    &\qquad\qquad+ \bar{\rho} \int \left[ \zeta_{r} \vec{\Omega}_F \cdot \frac{\partial \vec{\Omega}_F}{\partial \nabla \vec{v}_{\alpha \beta}}  \right]  \psi  \text{d}\uu,
\end{split}
\end{equation}
where 
\begin{subequations}
\begin{align}
    \vec{f}&=-\frac{\hat{\vec{R}}\tilde{U}}{\zeta_r+\zeta_F(F_N,\vec{v}_{c})}\\
    \vec{\Omega}_F&= -\frac{\zeta_F(F_N,\vec{v}_{c})}{\zeta_r}\vec{f}.
\end{align}
\end{subequations}
Note that the stress tensor $\boldsymbol{\sigma}^F$ depends self-consistently on the total stress tensor through our assumption that the normal force $\vec{F}_N(\boldsymbol{\sigma})$ depends on the total stress tensor.
\end{widetext}

\section{A Numerical Study} \label{sec:wsimu}
To examine the dynamical states predicted by our model, we use Eq.~\ref{full-omega-min} to evolve a set of $N$ rods' orientations until a steady state is established. We assume a simple shear flow geometry with
\begin{equation} \label{simple-shear-def}
    \vec{v} = \dot{\gamma} y \uvec{x},
\end{equation}
where $y$ is the height in the gradient direction. As we will only treat the rods' orientations, this means we lack spatial information and so will restrict to solving the fully solid frictional case, $p \gg 1$ (the lubricated case maps on to the Doi-Edwards theory with a modified drag constant, see Eq.~
\ref{eq:lub_contacts_drag}). For $\Theta(p) = 1$ in Eq.~\ref{full-omega-min} we have 
\begin{equation} \label{solidfric-omega-simu}
    \vec{\omega}^{i}_{\text{min}} =  \vec{\omega}_{0} - \frac{\Rhat\tilde{U}}{\zeta_{r} + \mu_{k} |\vec{F}_{N}| \left \langle  \frac{(\epsilon^{i})^2}{|\vec{v}_{c}(\vec{\omega}_{\textrm{min}})|} \right\rangle_{j} }.
\end{equation}
In Appendix~\ref{app:numerical-deriv}, we present a procedure to address the self-consistency within this expression by truncating a series expansion in $\vec{\omega}_{0}$ and applying the approximation of Appendix~\ref{appendix:v-c-approx}, resulting in
\begin{equation} \label{w-eq-solve}
\vec{\omega}^{i}_{\text{min}} = \vec{\omega}_{0} - D_{r}\frac{\Rhat\tilde{U} / k_{B}T}{1 + C(\phi, L/D, \mu_{k})\frac{[1 - \uu^{i}\uu^{i}:\tens{Q}]}{\left(\uu^{i}_x\uu^{i}_y\right)^2 + Q_{xy}^2}}.
\end{equation}

Here,
\begin{equation}
    C = \frac{3 \mu_k \phi L^2 |Q_{xy}| \left(\frac{\pi}{4(1-S^2)}\right)^{\frac{2}{3}}}{4\sqrt{2} D \left( 3 \pi L \phi^{\frac{1}{3}} - 8 \mu_k \left(\frac{\pi}{4(1-S^2)}\right)^{\frac{2}{3}} \text{Tr}(\tens{K}(\uu)) \right)},
\end{equation}
and the function $\text{Tr}(\tens{K}(\uu))$ (Eq.~\ref{trace-k}) constitutes an effective jamming metric: the rapid increase of this function drives the rotational mobility towards zero. A small parameter $\epsilon \sim .001$ is introduced to regularize Eq.~\ref{w-eq-solve} and prevent a singularity during numerical calculations; if the denominator of $C(\phi, L/D, \bar{\rho}, \mu_{k})$ approaches zero, then it is replaced by $\epsilon$. The quantity $\frac{[1 - \uu^{i}\uu^{i}:\tens{Q}]}{\left(\uu^{i}_x\uu^{i}_y\right)^2 + Q_{xy}^2}$ is always positive. This function scaled by $C$ represents the effective drag due to rod interactions, which also cannot be negative (see Eq.~\ref{tr-de-stress} in Appendix~\ref{app-w-eq-solve}). To ensure so, the regularization is performed so that $C$ approaches zero from the positive side alone. This choice reflects that the divergence of the normal force signifies a jammed state. Similarly, if the term proportional to $1/(1 - S^2)^{2/3}$ approaches divergence, it is replaced by $1/\epsilon$. The rotational diffusion constant $D_{r}$ defines  a natural timescale. Consequently, we can adjust the Peclet number $\textrm{Pe}=\dot{\gamma}/D_r$ by varying  $\dot{\gamma}$. The rod diameter $D$ is the natural length scale. 

\begin{figure*}[htb] \centering\includegraphics[width=.85\textwidth]{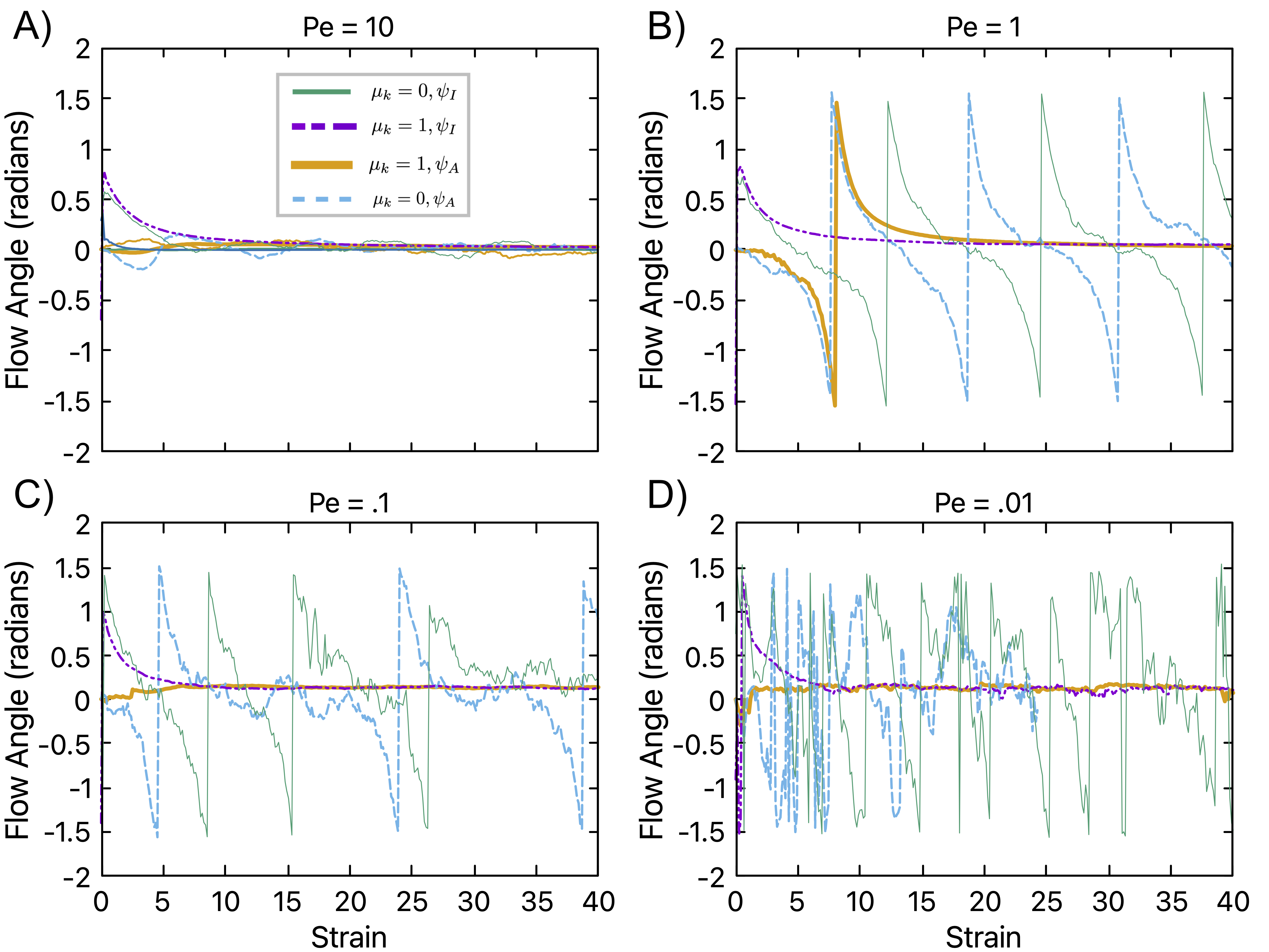} \label{fig:wsimu}
\caption{Results of solving Eq.~\ref{w-eq-solve} with and without friction, for varying Peclet number and two initial distributions: one, $\psi_{A}$, well aligned with the flow direction with $S\simeq0.8$; and one isotropic, $\psi_{I} \approx 1/(4 \pi)$. Here, $\phi = 0.43$, $L/D = 10$, $N=1000$. The thick orange curve is for $\mu_{k}=1$ and $\psi_{A}$, the purple dotted and dashed for $\mu_{k} =1$ and $\psi_{I}$, the dashed blue for $\mu_{k} = 0$ and $\psi_{A}$, and the thin green for $\mu_{k} = 0$ and $\psi_{I}$. A) Pe = 10, B) Pe = 1, C) Pe = .1, D) Pe = .01}. \label{fig:peclet_grid}
\end{figure*}

Eq.~\ref{w-eq-solve} presents essentially three pieces to numerically solve for a given rods' $\vec{\omega}_{\text{min}}$. First there are the two components of $\hat{R}\tilde{U}=  \hat{R}[k_{B}T\ln(\psi) + U(\uu)]$. We model the first term using  kicks applied to the rods' orientation \cite{ottinger1996}, which obeys the fluctuation dissipation theorem, and is of the form
\begin{equation}
\vec{\omega}^{i}_{\text{brown}} = \sqrt{\frac{2}{ \left(1 + \left(C\frac{[1 - \uu^{i}\uu^{i}:\tens{Q}]}{\left(\uu^{i}_x\uu^{i}_y\right)^2 + Q_{xy}^2}\right)/\zeta_{r}\right)\text{d}t}} \vec{\xi},
\end{equation}
where $\text{d}t$ is the size of the simulation timestep and $\vec{\xi}$ is a vector with each component independently drawn from a Gaussian distribution with zero mean and variance of unit. For the second term, we follow \cite{maier} and use the Maier-Saupe mean field potential,
\begin{equation}
    U(\uu^{i}) = - U_{MS} (\uu^{i} \cdot \tens{Q} \cdot \uu^{i}),
\end{equation}
which results in the torque 
\begin{equation}
    \vec{\Gamma}^{i}_{\text{MS}} = 2 U_{MS}\left(\uu^{i} \times \left[\tens{Q} \cdot \uu^{i} \right]\right),
\end{equation}
where $U_{MS}$ is the Maier-Saupe coupling constant. Following \cite{doiedwards} we set $U_{MS} = k_{B}T  (15/8)\phi L/D$. The third calculation is the denominator of the second term in Eq.~\ref{w-eq-solve} and $\omega_{0}$ (Eq.~\ref{flow-omega}), each of which can be constructed exactly at each step of the simulation from $\tens{Q}$ and the set of $\uu$.

\begin{figure*}[htb] \centering\includegraphics[width=.8\textwidth]{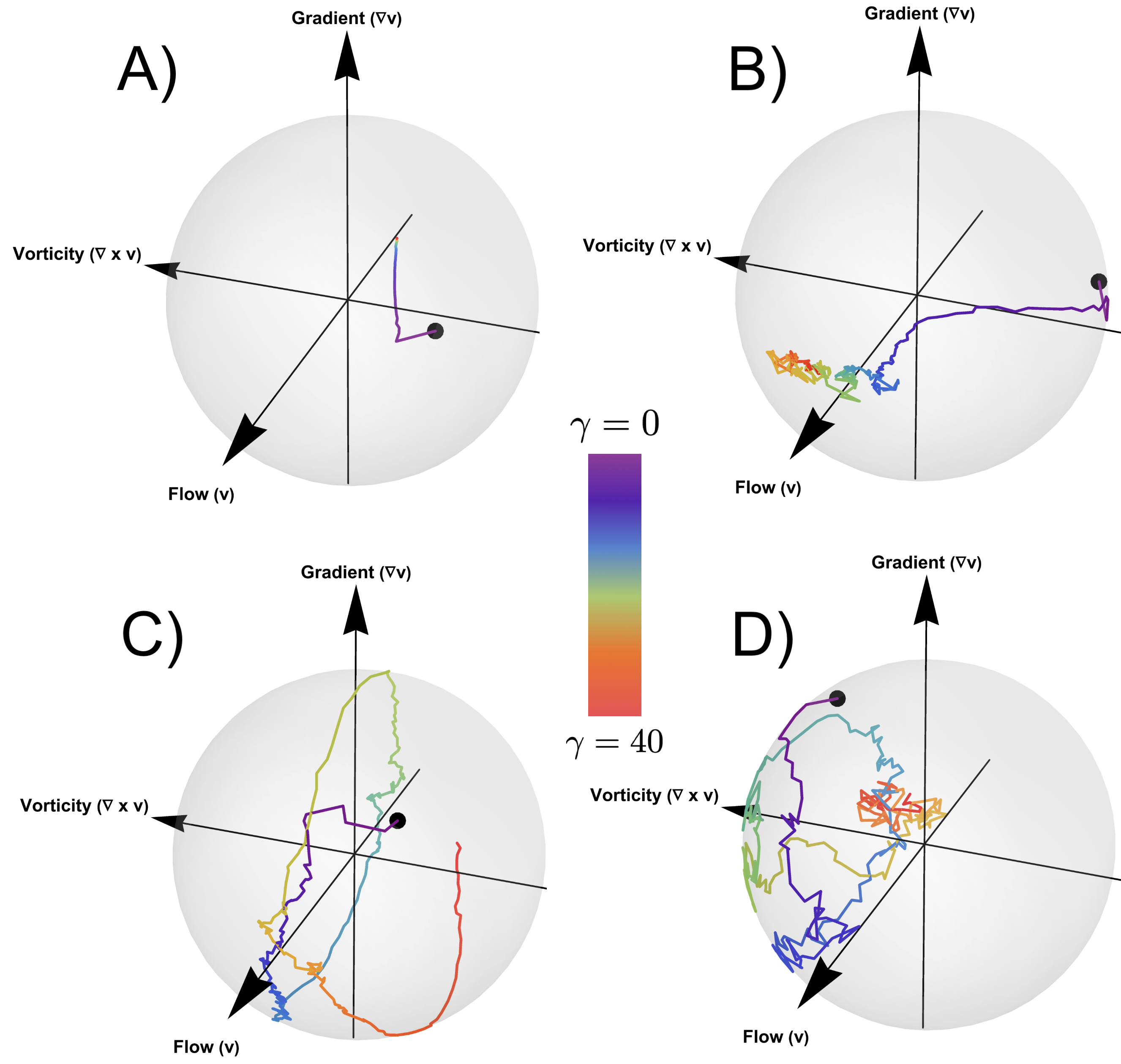}
\caption{Trajectories of the director plotted on a unit sphere for various curves of Fig.~\ref{fig:peclet_grid} with flow, gradient, and vorticity axes labeled. Black dot is initial position. Path coloring indicates value of strain ($\gamma$) in trajectory: dark purple is $\gamma \sim 0$, red is $\gamma \sim 40$. For all data plotted, $\phi = .43$, $L/D = 10$, and the initial condition was isotropic ($\psi_{I}$). A) $\text{Pe} = 10$, $\mu_{k} = 1$, B) $\text{Pe} = .01$, $\mu_{k} = 1$, C) $\text{Pe} = 1$, $\mu_{k} = 0$, D) $\text{Pe} = .1$, $\mu_{k} = 0$.} \label{fig:traj_grid}
\end{figure*}

This particle-based Langevin approach computes discrete rotational velocities for individual rods, and is a tractable alternative to solving the complex, unapproximated Smoluchowski equation. However, as the rotational drag coefficient in our model is explicitly orientation dependent, $\zeta_F = \zeta_{F}(\uu)$, the stochastic noise amplitude must also be orientation-dependent. To ensure our Langevin implementation maintains strict mathematical equivalence to the Smoluchowski equation, the numerical integration must be supplemented with a Stratonovich spurious drift correction \cite{ottinger1996}. This introduces an additional deterministic rotational velocity, $\omega_{\text{spur}}^i$, proportional to the orientational gradient of $1/(\zeta_{r} + \zeta_{F})$:
\begin{equation}
    \vec{\omega}_{\text{spur}} = -\frac{1}{(\zeta_{r} + \zeta_{F})^2} \left(\uu \times \left(\frac{\partial}{\partial \uu} \zeta_{F}\right)\right).
\end{equation}
This corrective drift repels the director from states of artificially high drag, preventing unphysical probability accumulation. With the total kinematics defined as
\begin{equation}
\omega_{\text{tot}}^i = \omega_0^i + \frac{\tau_{\text{MS}}^{i}}{\left[\zeta_r\left(1+C\frac{[1 - \uu^{i}\uu^{i}:\tens{Q}]}{\left(\uu^{i}_x\uu^{i}_y\right)^2 + Q_{xy}^2}\right)\right]} + \omega_{\text{brown}}^i + \omega_{\text{spur}}^i,
\end{equation}
we update each rods' orientation via
\begin{subequations}
\begin{align}
    \vec{u}^{i}(t+\text{dt}) & = \vec{u}^{i}(t) + \text{dt}\dot{\vec{u}}^{i}(t) \\
    & =  \vec{u}(t) + \text{dt}(\vec{\omega}^{i}_{\text{tot}}(t) \times \vec{u}^{i}(t)).
    \end{align}
\end{subequations}
To guarantee numerical stability and high resolution for our simulations, we choose $\text{dt}$ such that $\tau_{\dot{\gamma}}$ is significantly larger. Specifically, we maintain $\dot{\gamma}\text{dt} \lesssim 0.01$.

Our findings are collected in Fig.~\ref{fig:wsimu}, where for $\phi = 0.43, L/D = 10$ we plot the flow angle
\begin{equation}
    \theta = \arctan\left(\frac{n_{y}}{n_{x}} \right),
\end{equation}
where $n_{y}$ and $n_{x}$ are the components of the director (see Eq.~\ref{q-tens-def}) in the gradient and flow directions, respectively. We consider two possible initial conditions (randomly generated at $t=0$): an anisotropic $\psi(\uu,t=0)$ with $S \simeq .8$ and a director pointing in the flow direction, and an isotropic $\psi(\uu,t=0)$ with $S\sim0$.

Without friction ($\mu_{k} = 0$), we are simulating the Doi model (Eq.~\ref{solidfric-omega-simu} becomes exactly Eq.~\ref{doi-rayl-flow-omega}), and our findings match the well-known dynamical regimes for this model \cite{larson_arrested}. At high shear rates ($\text{Pe} = 10$, Fig.~\ref{fig:peclet_grid}A), strong advective flow arrests continuous rotation; the director is forced into a steady, flow-aligned state at a small positive angle $\theta$, regardless of whether the system starts from an initially aligned ($\psi_A$) or isotropic ($\psi_I$) configuration. As the Peclet number is decreased to $\text{Pe} = 1$ (Fig.~\ref{fig:peclet_grid}B), this flow-aligned state loses stability. Both initial conditions transition into a periodic ``kayaking" orbit, where $\theta$ continuously shifts from $-\pi/2$ to $+\pi/2$ while the director maintains a significant tilt along the vorticity axis, see Fig.~\ref{fig:traj_grid}C. These dynamics match the bifurcation analysis of \citet{faraoni}, who demonstrated that orbits in the flow-gradient plane are unstable to out-of-plane perturbations—here provided by Brownian fluctuations. Further decreasing the Peclet number ($\text{Pe} \le 0.1$, Fig.~\ref{fig:peclet_grid}C,D) allows thermal motion to dominate the orientational dynamics. The system transitions from a well-defined kayaking orbit into noisy, erratic out-of-plane orientational wandering. See Fig.~\ref{fig:traj_grid}D for an example of such.

With friction ($\mu_{k}=1$), we are outside established theory. The general effect presented in Fig.~\ref{fig:peclet_grid} is that motion in the flow-gradient plane is suppressed as compared to the frictionless case. While $\theta$ may still pass from $-\pi/2$ to $+\pi/2$ and the director kayaks (such as in Fig.~\ref{fig:peclet_grid}B), these events are sporadic and are a product of the initial director, which is due to the initially imposed $\psi$, having a negative gradient component. Aside from these events, for all Pe and initial conditions, the director stays for long periods in a flow-aligned state. See  Fig.~\ref{fig:traj_grid}A for an example trajectory that flow aligns quickly despite a $\psi_{I}$ initial condition. For smaller $\text{Pe} = .1, .01$, the vorticity component of the director in the flow-aligned state drifts about 0, with larger magnitude for an initial condition of $\psi_{I}$ (as shown in Fig.~\ref{fig:traj_grid}B).

\section{Conclusion}

We have shown how dissipation due to contact friction can be incorporated into the Rayleighian of rods in suspension of rigid rods under flow, in terms of the  angular velocity $\vec{\omega}$ for rigid rod rotation.
We apply  Onsager’s variational method  \cite{doionsager} and minimize the Rayleighian with respect
to $\vec{\omega}$, and thus derive a new Smoluchowski equation and stress tensor $\boldsymbol{\sigma}$ for flowing rigid rod suspensions.  If the contribution to the torque on a given rod from frictional contacts is weak (compared to that from viscous forces), this Smoluchowski equation leads to a dynamical equation for the order parameter tensor $\tens{Q}$ analogous to the Doi model \cite{doiedwards}, with an additional  term  proportional to the magnitude of the average torque due to frictional contacts. Correspondingly, the stress tensor is that of the Doi model with additional terms,
due to frictional contacts, which scale with the magnitude of the average frictional torque. We identify distinct contributions from lubricated or solid contacts.

The expressions for the $\tens{Q}$ dynamics and stress involve an average over the  reciprocal of the contact velocity between rods, which impedes further analytic progress. The expressions for $\dot{\tens{Q}}$ (Eq.~\ref{gen-Q-dot-with-fric}) and $\boldsymbol{\sigma}$ (Eq.~\ref{sig_main_fric}) can, in principle, be simplified by approximating the distributions of contact velocities (Appendix \ref{appendix:v-c-approx}). A numerical approach could also be also adopted to address a given flow geometry (such as planar shear) as done in \cite{yamane}.

Critical to our approach is treating frictional dissipation (due to two-particle interactions) via a mean field, \textit{i.e.} by averaging over the distribution of the relevant positional and orientational variables of the interacting pairs. In deriving this mean field we obtained an equation (Eq.~\ref{final-c-s}) for the mean number of contacts per rod $\langle c \rangle$, as a function of the scalar nematic order parameter $S$ and nematic strength $\phi L/D$. This form extends the known isotropic result \cite{randomcontact} to the ordered regime, such as that found for dense suspensions of rods under shear. Verification of this function is challenging experimentally \cite{dbonn}, but would be straightforward in simulations. 

In our formulation, if there are solid frictional contacts we find a coupled, self-consistent set of equations for the dynamics of the order tensor $\tens{Q}$, the stress tensor $\boldsymbol{\sigma}$, and the microscopic dynamics $\vec{\omega}^{i}_{\text{min}}$ of the rods.
The frictional contribution to the stress tensor $\boldsymbol{\sigma}$ depends on the normal force $|F_{N}(\boldsymbol{\sigma})|$ between particles, which is controlled by the fluid stress tensor itself, and hence the shear rate $\dot{\gamma}$, leading to a self consistent relation to determine the stress. The rod rotation rate $\vec{\omega}^{i}_{\text{min}}$ and hence the order tensor $\boldsymbol{\sigma}$ are controlled by the contact velocity $\vec{v}_{c}$ (see Eq.~\ref{full-omega-min}), which is, by Eq.~\ref{full-rel-v}, a function of $\vec{\omega}^{i}_{\text{min}}$ and the force $|F_N(\boldsymbol{\sigma})|$. The self-consistent nature of our equation set can be traced back to the form of Coulomb friction, which is a non-polynomial function of $\vec{\omega}$. This leads to a balance of torques that is non-linear in $\vec{\omega}$ (\textit{i.e.} a necessarily self-consistent $\vec{\omega}^{i}_{\text{min}}$). In Sec.~\ref{sec:wsimu} we develop an  approximation to handle the self-consistency in the equation set by a truncated series expansion( Appendix~\ref{appendix:v-c-approx}) and an approximation to the normal force in terms of the average pressure (Eq.~\ref{press-def}). Interestingly, the inclusion of frictional torques in the dynamics of dense rods under shear results in suppression of motion of the director in the flow-gradient plane.

In the limit of only lubricated contacts, \emph{i.e.} $\Theta(p) = 0$ in Eq.~\ref{full-omega-min}, we find an increase to the rotational drag constant in the $\tens{Q}$ equation. 
Importantly, this perturbation to the drag constant ($\tilde{\zeta}_{r} - \zeta_{r}$ in Eq.~\ref{eq:lub_contacts_drag}) is not simply proportional to the average number of contacts in the isotropic case, $\phi L/D$. This is because  Eq.~\ref{BL}, which defines the form of the boundary lubricated  between rods, includes a coefficient of $1/|\vec{\tau}|=1/|\uu^i\times\uu^j|$, 
which cancels the excluded volume term (proportional to $\vec{\tau}$) eventually applied in  Eqs.~\ref{full-avg-lub},~\ref{torque-mf-lub} to count the expected number of contacts.

In previous work, inclusion of frictional contacts in rod dynamical models was either limited to boundary lubricated friction (Eq.~\ref{BL}), as in \cite{tollmanson,lubefric}, or was found to have no effect on $\dot{\tens{Q}}$ \cite{bounoua,sandstrom}. The latter finding was due to the assumption that the frictional torque between rods could be calculated directly by averaging over the expected distribution of interacting rods, treating each as undergoing the motion prescribed by Jeffery's equation, $\dot{\vec{u}}_{\text{Jeffery}}$. This contact torque would then be included as a term in the equation of motion for the rod motion, $\dot{\vec{u}}$. In \cite{bounoua,sandstrom} the average of the torque over contact positions results in a zero mean contact torque. By contrast, the frictional contact dissipation function that we construct is positive semidefinite in the relative rod-rod contact velocity $|\vec{v}_{c}|$, and we determine $\dot{\vec{u}}$ from the balance of torques produced from minimizing the associated Rayleighian, which leads to a non-zero contact torque.

A fundamental assumption in our approach is that solid frictional contacts are undergoing sliding friction and have exceeded the static limit. This allows us to include the power dissipated during friction in the Rayleighian as a function of $\vec{\omega}$. However, in reality the rods will stick and interact through solid friction without relative motion; this effect cannot be treated by a variational approach, as static friction does no work. Large clusters of stuck rods can contribute to shear thickening \cite{heussinger}, so our model misses this contribution. A potential solution might be to consider the dissipation of larger structures (such as clusters of rods), but that is beyond the scope of this project. It has been shown in experiments on dense sphere suspensions \cite{guazzelli_influence} that the maximum flowable volume fraction decreases with increasing roughness. As we always allow our rods to slide past each other, this behavior is naturally outside of the model's scope. We have also assumed that suspensions are spatially homogeneous, which may break down during DST. It has recently been shown that in the regime of DST there are significant stress fluctuations in tandem with nonaffine flow \cite{esmaeel}. The observation that frictional forces can induce branched frictional contact networks \cite{singhcontacts} also suggests a more careful treatment that includes spatial inhomogeneity.

\acknowledgments

We thank Lucas Cunha, Emanuela Del Gado, Mauro Mugnai, Vikram Rathee, and Jeff Urbach for helpful discussion and  correspondence. We are grateful to Georgetown University and the Ives Foundation for support.

\section*{Data Availability}

Data sharing is not applicable to this article as no new data were created or analyzed in this study.

\clearpage

\appendix

\counterwithin{equation}{section}
\renewcommand{\theequation}{\thesection.\arabic{equation}}

\begin{widetext}
    
\section*{Nomenclature}
Nomenclature used, including symbol, definition, and if relevant, either a definition or the defining equation. In this paper, superscript $i$ and $j$ denotes particle index. Subscript Greek indices denote Einstein notation, with implied summation over repeated indices. Lowercase bold symbols denote vectors; uppercase bold symbols denote tensors and operators.

\setcounter{table}{0} 
\begin{table}[h]
\refstepcounter{table} 
\label{table:nomen}
\begin{center}
\begin{tabular}{ ccc }
 L   &  Length of rod   &   \\
 D   &  Diameter of rod   &   \\
$\phi$   &  Volume fraction   & $N v_{\text{rod}}/V$  \\
$\nabla v$   &  Velocity gradient   & Eq.~\ref{veloc-grad-def}  \\
$\tens{D}$   &  Symmetric velocity gradient   & $(1/2) \left(\nabla v + \nabla v^{T} \right)$  \\
$\eta_{s}$   &  Viscosity of solvent &   \\
$\uu$   &  Rod orientation  &   \\
$\epsilon$ & Point of contact along rod & Eq.~\ref{epsilon} \\
$\vec{\omega}$   &  Rod angular velocity  &   \\
$\vec{\tau}$   &  Vector normal to plane defined by rods  & $\uu^{i} \times \uu^{j}$  \\
$|\vec{F}_{N}|$   &  Normal force between rods  &  \\
$\mu_{k}$   &  Sliding frictional coefficient & \\
$\hat{R}$   &  Rotational differentiation operator  & Eq.~\ref{rot-operator}  \\
$\psi$   &  Orientational distribution function &   \\
$\tens{Q}$   &  Orientational tensor   & Eq.~\ref{q-tens-def}  \\
$S$   &  Order parameter   & Eq.~\ref{q-tens-def}  \\
$\mathcal{R}$   &  Rayleighian function  & Eq.~\ref{rayl-def}  \\
$\Phi$   &  Dissipation function   &   \\
$\tilde{U}$   &  Brownian and Excluded volume potential  & $\kbT \ln{\psi} + U\left(\hat{\vec{u}} \right)$  \\
$\zeta_{r}$   &  Dilute rotational drag constant  &   \\
$\zeta_{F}$  &  Rotational drag in the presence of friction  & ~Eq.~\ref{eq:zeta_f} \\
$\Delta$  &  Dimensionless ratio of frictional to shear torque  & ~Eq.~\ref{delta} \\
$\Theta$  & Switching function between solid and lubricated  & ~Eq.~\ref{contact-split} \\[3truept]
$p$  &  Ratio between pressure and characteristic pressure & $\dfrac{P}{P^{*}}$ \\[6truept]
$\boldsymbol{\sigma}$  &  Stress tensor  & ~Eq.~\ref{sig_main_fric} \\
\end{tabular}
\end{center}
\end{table}
\end{widetext}

\section{Average ``Fiber" Torque}
\label{appendix:fiber-torque}
An alternative method to determine the effect of contact torque in suspensions of rods was explored by \citet{bounoua}, \citet{djalili}, \citet{tollmanson}, and \citet{sandstrom}. In this approach the angular equation of motion $\dot{\vec{u}}$ is a sum of the advective dynamics given by  Jeffery's Equation \cite{jeff}, and the torque $\vec{\Gamma}$ due to frictional contacts:
\begin{equation}
    \dot{\vec{u}} = \dot{\vec{u}}_{\text{Jeffery}} + \alpha \langle \vec{\Gamma}(\vec{v}_{c}) \rangle_{j}.
\end{equation}
Here, $\alpha\sim1/\zeta_r$ is a rotational drag coefficient proportional to the magnitude of the torque part of the resistance tensor \cite{kim-resist}, the angle brackets are the same as in Eq.~\ref{mean-field}, and
\begin{equation} \label{u-dot-jeff}
    \dot{\vec{u}}_{\text{Jeffery}} = \tens{\Omega} \cdot \uu^{i} + \lambda \left[\tens{D} \cdot \uu^{i} - \left(\uu^{i} \cdot \tens{D} \cdot \uu^{i}\right) \uu^{i} \right],
\end{equation}
where $\lambda$ depends only on rod aspect ratio \cite{larsonbook},
\begin{equation}
    \lambda = \frac{\left(\frac{L}{D}\right)^2-1}{\left(\frac{L}{D}\right)^2+1}.
\end{equation}
As $\vec{\Gamma}$ depends on $\vec{v}_{c}$, it implicitly depends on $\dot{\vec{u}}$, according to
\begin{equation} \label{pre-jeff-v-c}
    \vec{v}_{c}^{ij} = \epsilon^{i} \dot{\vec{u}}^{i} - \epsilon^{j} \dot{\vec{u}}^{j} + \nabla \vec{v} \cdot (\vec{r}_{\text{com}}^{i} - \vec{r}_{\text{com}}^{j}),
\end{equation}
which is equivalent to Eq.~\ref{v-c} (recall that $\vec{\omega} \times \uu =\dot{\vec{u}}$). To resolve this \cite{bounoua,djalili,tollmanson,sandstrom}, one can replace $\dot{\vec{u}}$ in the contact velocity by the dilute rod form due to Jeffery (Eq.~\ref{u-dot-jeff}), to write $\vec{v}_c$ explicitly in $\uu^{i},\uu^{j},\epsilon^{i},\epsilon^{j}$, and $\nabla \vec{v}$. If it is assumed that rods are in close proximity the term linear in $\nabla \vec{v}$ can be dropped and in the limit $\lambda \simeq 1$, (\textit{i.e.} $\frac{L}{D} \gg 1$), the contact velocity becomes
\begin{equation} \label{eq:old-vc-approx}
    \vec{v}_{c} \simeq  \epsilon^{j} (\uu^{j} \cdot \tens{D} \cdot \uu^{j}) \uu^{j} - \epsilon^{i} (\uu^{i} \cdot \tens{D} \cdot \uu^{i}) \uu^{i}.
\end{equation}
This form was used by \citet{sandstrom} and \citet{djalili} to address the boundary lubricated case.  \protect{\citet{bounoua}} used \ref{eq:old-vc-approx} in their treatment of solid frictional torque to obtain 
\begin{subequations}
\begin{align}\label{eq:torquebounouma}
    \vec{\Gamma} & = \alpha \mu_{k} |F_{\text{N}}| \left\langle \frac{\hat{u}_{i} \times (\epsilon_{i} \hat{u}_{i} \times \hat{v}_{c})}{|\vec{v}_{c}|} \right\rangle_{j} \\
           & = \alpha \mu_{k} |F_{\text{N}}| \left\langle \frac{\epsilon^{i} \epsilon^{j}}{|\vec{v}_{c}|} \left(\uu^{j} \cdot \tens{D} \cdot \uu^{j}\right) \left[\uu^{i}(\uu^{i} \cdot \uu^{j}) - \uu^{j} \right]\right\rangle_{j}.
\end{align}
\end{subequations}
This torque vanishes upon averaging over the contact positions, and $\dot{\vec{u}}$ is unchanged. However, if one uses $\dot{\vec{u}}_{\text{Jeffery}}$ in Eq.~\ref{pre-jeff-v-c} and makes no further assumptions about $\vec{v}_{c}$, the contribution is instead
\begin{multline}
    \vec{\Gamma} = \mu_{k} |\vec{F}_{\text{N}}| (1 - \lambda) \left\langle \frac{(\epsilon^{i})^{2}}{|\vec{v}_{c}|}\left[ \tens{D} \cdot \uu^{i} - (\uu^{i} \cdot \tens{D} \cdot \uu^{i})\uu^{i} \right]\right. \\
    \left.\phantom{\dfrac{10^2}{10}}+ \left.\mathcal{O}\left(\epsilon^{i} \epsilon^{j} \right) \right]\right\rangle_{j}.
\end{multline}
The terms proportional to $\epsilon^{i} \epsilon^{j}$ will average to zero as before, but the first term is non-zero for $\lambda \neq 1$.

\begin{widetext}
\section{Stress Tensor}
\label{appendix:doi-stress} 
The stress tensor for our model (Eq.~\ref{eq:stress-total}) has  two parts: that due to dissipation from non-frictional sources, and a contribution from frictional dissipation (both lubricated and solid). The former is modified by the minimum angular velocity of the rods, and thus in the following we insert Eq.~\ref{full-omega-min} and collect terms to show the relation to the Doi model. The latter frictional born terms we also calculate here. 
From Eq.~\ref{eq:stress-total}, the non-frictional dissipation term is $\bar{\rho}\frac{\delta \mathcal{R}_{0}}{\delta (\nabla v)_{\alpha \beta}}$; so we have 
 \begin{align}
     \boldsymbol{\sigma}^{0}_{\alpha \beta} & = \bar{\rho}\frac{\partial \mathcal{R}_{0}}{\partial (\nabla v)_{\alpha \beta}} \label{eq:sigma0}\\
     & = \bar{\rho} \int \left[  \zeta_{r} (\vec{\omega} - \vec{\omega}_{0})\cdot \frac{\partial (\vec{\omega}-\vec{\omega}_0)}{\partial (\nabla \vec{v})_{\alpha \beta}} + \frac{\zeta_{r}}{2} (\uu \cdot \nabla \vec{v} \cdot \uu) u_{\alpha} u_{\beta} 
     + \frac{\partial \vec{\omega}}{\partial (\nabla\vec{v})_{\alpha \beta}} \cdot \hat{R}\tilde{U} \right] \psi  \text{d}\uu + 2 \eta_{s} \tens{D}. \label{eq:sigma0full}
\end{align}
 The first two terms in brackets constitute the form of the stress tensor for non-interacting frictional suspensions derived by  \citet{doibook} using the Rayleighian approach, while the last term due to $U(\uu)$ appears in the presence of excluded volume interactions, as derived in \cite{doiedwards}. The angular velocity $\vec{\omega}$ is given by Eq.~\ref{full-omega-min}, and can be written as 
\begin{align} \label{eq:omega-full}
\vec{\omega}&=\vec{\omega_{\textrm{DE}}}+ \frac{\zeta_F(F_N,\vec{v})}{\zeta_r+\zeta_F(F_N,\vec{v})}\frac{\hat{\vec{R}}\tilde{U}}{\zeta_r}, 
\end{align}
where 
\begin{align}
\vec{\omega_{\textrm{DE}}}&=\vec{\omega}_0 - \frac{\hat{\vec{R}}\tilde{U}}{\zeta_r} \label{eq:omega_DE}
\end{align}
is the angular velocity used by Doi and Edwards \cite{doiedwards}. 

In the Doi-Edwards case, for $\vec{\omega}=\vec{\omega}_{\textrm{DE}}$, the stress tensor $\boldsymbol{\sigma}^{0,\textrm{DE}}_{\alpha \beta}$ is
\begin{equation}\boldsymbol{\sigma}^{0,\textrm{DE}}_{\alpha \beta} = \bar{\rho} \int \left[  \zeta_{r} (\vec{\omega}_{\textrm{DE}} - \vec{\omega}_{0})\cdot \frac{\partial (\vec{\omega}_{\textrm{DE}} - \vec{\omega}_{0})}{\partial (\nabla \vec{v})_{\alpha \beta}} + \frac{\zeta_{r}}{2} (\uu \cdot (\nabla \vec{v}) \cdot \uu) u_{\alpha} u_{\beta} 
     + \frac{\partial \vec{\omega}_{\textrm{DE}}}{\partial (\nabla \vec{v})_{\alpha \beta}} \cdot \hat{R}\tilde{U} \right] \psi  \text{d}\uu + 2 \eta_{s} \tens{D}.
\end{equation}
Using $\frac{\partial \vec{\omega}_{\textrm{DE}}}{\partial(\nabla \vec{v})_{\alpha \beta}} = \frac{\partial \vec{\omega}_{0}}{\partial (\nabla \vec{v})_{\alpha \beta}}$, which follows from Eq.~\ref{eq:omega_DE},
 this simplifies to
\begin{equation} \label{simplified-de-stress}
    \boldsymbol{\sigma}^{0,\textrm{DE}}_{\alpha \beta} = \bar{\rho} \int \left[\frac{\zeta_{r}}{2} (\uu \cdot \nabla \vec{v} \cdot \uu) u_{\alpha} u_{\beta} 
     + \frac{\partial \vec{\omega}_{\textrm{DE}}}{\partial (\nabla \vec{v})_{\alpha \beta}} \cdot \hat{R}\tilde{U} \right] \psi  \text{d}\uu + 2 \eta_{s} \tens{D}.  
\end{equation}
Using Eq.~\ref{eq:omega-full} in Eq.~\ref{eq:sigma0full} and comparing to Eq.~\ref{simplified-de-stress}, we find the frictional contribution to the angular rotation,
\begin{equation} 
    \boldsymbol{\sigma}^{0}_{\alpha \beta} = \boldsymbol{\sigma}^{0,\textrm{DE}}_{\alpha \beta} + \bar{\rho} \int \left[ \zeta_{r} \vec{\Omega}_F \cdot \frac{\partial \vec{\Omega}_F}{\partial (\nabla \vec{v})_{\alpha \beta}}  \right]  \psi  \text{d}\uu,
\end{equation}
where
\begin{equation}
    \vec{\Omega}_F= \frac{\zeta_F(F_N,\vec{v})}{\zeta_r+\zeta_F(F_N,\vec{v})}\frac{\hat{\vec{R}}\tilde{U}}{\zeta_r}.
\end{equation}

The second two terms in the stress tensor (Eq.~\ref{eq:stress-total}) account for contact dissipation.
If all frictional contacts are solid-like ($\Theta(p) = 1$), then by using Eq.~\ref{full-avg-solid} we find
\begin{subequations} \label{stress-addition-solid}
\begin{align}
    \frac{\boldsymbol{\sigma}_{\alpha \beta}^{\text{solid}}}{\bar{\rho}} & = \frac{\partial \Phi_{\text{sol}}^{\text{mf}}}{\partial (\nabla v)_{\alpha \beta}} \\
    & = \mu_{k} |\vec{F}_{N}| \left\langle  \left\langle  \frac{1}{|\vec{v}_{c}|} \vec{v}_{c,\mu} \frac{\partial \vec{v}_{c,\mu}}{\partial (\nabla v)_{\alpha \beta}} \right\rangle \right\rangle\\
    &  = \mu_{k} |\vec{F}_{N}| \left\langle  \left\langle  \frac{\vec{v}_{c,\alpha}}{|\vec{v}_{c}|} \left( \epsilon^{j} \uu^{j} - \epsilon^{i} \uu^{i} \right)_{\beta} + \frac{2(\epsilon^{i}\uu^{i} \times \vec{v}_{c})}{|\vec{v}_{c}|} \cdot \frac{\partial \vec{\omega}^{i}}{\partial (\nabla v)_{\alpha \beta}} \right\rangle \right\rangle \\ 
    & = 2 \mu_{k} |\vec{F}_{N}| \left\langle  \left\langle \frac{(\epsilon^{i})^2}{|\vec{v}_{c}|} \left[ -(\vec{\omega}^{i} \times \uu^{i})_{\alpha} \uu^{i}_{\beta} + (\nabla \vec{v} \cdot \uu^{i})_{\alpha} \uu^{i}_{\beta}  +  (\vec{\omega}^{i} - \vec{\omega}^{i}_{0}) \cdot \frac{\partial \vec{\omega}^{i}}{\partial (\nabla v)_{\alpha \beta}} \right]\right\rangle \right\rangle \\
    & = 2 \mu_{k} |\vec{F}_{N}| \left\langle \left\langle \frac{(\epsilon^{i})^2}{|\vec{v}_{c}|} \, \left[ \uu^{i}_{\alpha}\uu^{i}_{\beta}\,\left(\uu^{i}\!\cdot\!\vec{\nabla} \vec{v}\!\cdot\!\uu^{i}\right)
    +\vec{f} \cdot \frac{\partial \vec{f}}{\partial (\nabla v)_{\alpha \beta}} \right]\right\rangle \right\rangle.
\end{align}
\end{subequations}
where we have used Eq.~\ref{full-rel-v} in the third and fourth lines. We have  used Eq.~\ref{full-omega-min} to write $\vec{\omega}^i\equiv\vec{\omega}_0 + \vec{f}$ in the final line, where 
 the rotational velocity $\vec{f}$ due to interactions is
\begin{equation}
    \vec{f}=-\frac{\hat{\vec{R}}\tilde{U}}{\zeta_r+\zeta_F(F_N,\vec{v})}.
\end{equation}
Note that $\zeta_r\vec{\Omega}^F=-\zeta_F(F_N,\vec{v}) \vec{f}$. Similarly, if all contacts are lubricated ($\Theta(p) = 1$), we use  Eq.~\ref{full-avg-lub} to find
\begin{subequations}
\begin{align}
    \frac{\boldsymbol{\sigma}_{\alpha \beta}^{\text{BL}}}{\bar{\rho}} & = \frac{\partial \Phi_{\text{BL}}^{\text{mf}}}{\partial (\nabla v)_{\alpha \beta}} \\
    & =  \left\langle  \left\langle \frac{2\eta_{s} D}{|\vec{\tau}|}  \vec{v}_{c,\mu} \frac{\partial \vec{v}_{c,\mu}}{\partial (\nabla v)_{\alpha \beta}}  \right\rangle \right\rangle \\
    & = 4  \eta_{s} D \left\langle\left\langle \frac{(\epsilon^{i})^2}{|\vec{\tau}|} \,\left[ \uu^{i}_{\alpha}\uu^{i}_{\beta}\,\left(\uu^{i}\!\cdot\!\vec{\nabla} \vec{v}\!\cdot\!\uu^{i}\right)
    +\vec{f} \cdot \frac{\partial \vec{f}}{\partial (\nabla v)_{\alpha \beta}} \right]\right\rangle \right\rangle \\
    & = 4  \eta_{s} D \left\langle\left\langle \frac{(\epsilon^{i})^2}{|\vec{\tau}|} \,\left[ \uu^{i}_{\alpha}\uu^{i}_{\beta}\,\left(\uu^{i}\!\cdot\!\vec{\nabla} \vec{v}\!\cdot\!\uu^{i}\right) \right]\right\rangle \right\rangle \\
    & = \frac{4 \pi \phi \eta_{s} L^3}{3} \int  \uu_{\alpha}\uu_{\beta}\,\left(\uu\!\cdot\!\vec{\nabla} \vec{v}\!\cdot\!\uu\right) \psi(\uu)  \text{d}\uu,
\end{align}
\end{subequations}
where in the fourth line we have used the independence of $f$  from $(\nabla v)_{\alpha \beta}$, and in the fifth line applied the definition of the angle brackets, Eq.~\ref{mean-field}. This form is that of the Doi-Edwards model (Eq.~\ref{simplified-de-stress}), with a modified leading coefficient and without the free energy driven term.
Both forms of friction contribute to  the anisotropic Newtonian stress tensor (first term) and have a higher order contribution that is second order in the rod-rod interactions.

\section{Approximate form of\,\,\, $\left\langle \dfrac{\left(\epsilon^{i} \right)^2}{|v_{c}|}\right\rangle_{j}$}
\label{appendix:v-c-approx}
In order to approximate $\left\langle  \frac{\left(\epsilon^{i} \right)^2}{|\vec{v}_{c}(\vec{\omega}_{\text{min}})|} \right\rangle_{j}$, which appears in Eqs.~\ref{full-omega-min} and \ref{gen-Q-dot-with-fric}, we must first consider $1/|{\vec{v}}_{c}(\vec{\omega}_{\text{min}})|$. We seek a form that removes the dependence on $\uu_{j}$, $\vec{\omega}_{j}$ , and $\vec{\omega}_{\text{min}}$. So, we assume $\vec{\omega}_{\textrm{min}}\approx\omega_0$, and make use of Eq.~\ref{flow-omega},

\begin{subequations} \label{omega-approx}
\begin{align}
    \vec{\omega}^{i}_{\text{min}} \simeq \vec{\omega}_{0}^{i} & = \uu^{i} \times (\nabla \vec{v} \cdot \uu^{i}), \\
    \vec{\omega}^{j}_{0} & = \uu^{j} \times (\nabla \vec{v} \cdot \uu^{j}).
\end{align}
\end{subequations}

We also use an approximation from \citet{bounoua}, 
\begin{equation} \label{ferec-approx}
    \frac{1}{|\vec{v}_{c}|} \approx \frac{|\tens{D} \epsilon^{i}|}{\left[ v_{c}^{2} \right]},
\end{equation}
where the square brackets denote the average
\begin{equation} \label{approx-angle-brackets}
    [x] \equiv \frac{1}{L^2}\int_{-\frac{L}{2}}^{\frac{L}{2}}{\text{d}\epsilon^{i}}\int_{-\frac{L}{2}}^{\frac{L}{2}}\text{d}\epsilon^{j} \int{ x \psi^{j} \text{d}\hat{\vec{u}}^{j}}.
\end{equation}
Ref.~\cite{bounoua} argues that the approximation above underestimates $1/|\vec{v}_c|$ by a factor of $\frac{1}{\ln{L/D}}$.
Inserting Eq.~\ref{omega-approx} into the contact velocity, Eq.~\ref{full-rel-v}, we find
\begin{equation}
\vec{v}_{c}^{2} = (\epsilon^{i})^2 (\uu^{i} \cdot \nabla \vec{v} \cdot \uu^{i})^{2} + (\epsilon^{j})^2 (\uu^{j} \cdot \nabla \vec{v} \cdot \uu^{j})^{2}.
\end{equation}
Using this form in Eq.~\ref{ferec-approx} gives
\begin{equation}
\begin{split}
\frac{1}{|\vec{v}_{c}|} & = \frac{|\tens{D} \epsilon^{i}|}{[ (\epsilon^{i})^2 (\uu^{i} \cdot \nabla \vec{v} \cdot \uu^{i})^{2} + (\epsilon^{j})^2 (\uu^{j} \cdot \nabla \vec{v} \cdot \uu^{j})^{2}]} \\[5truept]
& = \frac{12 |\tens{D} \epsilon^{i}|}{ L^{2} \left[\left(\uu^{i} \cdot \nabla \vec{v} \cdot \uu^{i}\right)^{2} + \left(\nabla v\right)_{\alpha \beta} \left(\nabla v\right)_{\mu \nu} \int u^{j}_{\alpha} u^{j}_{\beta} u^{j}_{\mu} u^{j}_{\nu} \psi^{j} \text{d}\uu^{j}\right] }, \\
\end{split}
\end{equation}
where in the second line we have used the definition of the square brackets (Eq.~\ref{approx-angle-brackets}). With an approximation for $\frac{1}{|\vec{v}_{c}|}$ we can now address the term that appears in $\dot{\tens{Q}}$ and $\boldsymbol{\sigma}$:
\begin{equation} \label{unsimplified-anglej}
\begin{split}
    \left\langle  \frac{\left(\epsilon^{i} \right)^2}{|\vec{v}_{c}(\vec{\omega}^{i}_{\text{min}})|} \right\rangle_{j} & \approx \frac{12|\tens{D}|}{L^{2}} \left\langle  \frac{ |\epsilon^{i}| \left(\epsilon^{i} \right)^2}{\left((\uu^{i} \cdot \nabla \vec{v} \cdot \uu^{i})^{2} + (\nabla v)_{\alpha \beta} (\nabla v)_{\mu \nu} \int u^{k}_{\alpha} u^{k}_{\beta} u^{k}_{\mu} u^{k}_{\nu} \psi^{k} \text{d}\uu^{k}\right)} \right\rangle_{j} \\
\end{split}
\end{equation}
The second term in the  denominator is an orientational average over four powers of  $\uu$, which we  cast in terms of $\tens{Q}$ by invoking  Doi's quadratic closure \cite{doimain},
\begin{equation} \label{quad_close}
    X_{\mu \eta} \langle u_{\alpha} u_{\beta} u_{\mu} u_{\eta} \rangle \approx X_{\mu \eta} \langle u_{\mu} u_{\eta} \rangle \langle u_{\alpha} u_{\beta} \rangle.
\end{equation}
While there exist many other potential closures \cite{corona}, some of which have been shown to be more accurate in shear flow, testing their validity in this context is outside our scope and here we only invoke Eq.~\ref{quad_close}.
By inserting this closure and assuming that the velocity gradient tensor is traceless (\textit{i.e.}, an incompressible fluid), we obtain
\begin{equation} \label{4thclosed}
    (\nabla v)_{\alpha \beta} (\nabla v)_{\mu \nu} \int u^{j}_{\alpha} u^{j}_{\beta} u^{j}_{\mu} u^{j}_{\nu} \psi^{j} \text{d}\uu^{j} \simeq (\nabla \vec{v}_{\alpha \beta} \tens{Q}_{\alpha \beta})^2.
\end{equation}
The angle brackets in \ref{unsimplified-anglej} include an integral over $|\vec{\tau}|\psi^{j}\text{d}\uu^{j}$ (see Eq.~\ref{mean-field}). Similar to that of \citet{doimain}, we can express $\vec{\tau}$ in terms of irreducible tensors built from $\uu^{j}$ and $\uu^{i}$,
\begin{equation}  \label{doi-tau}
\begin{split}
\int |\vec{\tau}|\psi^{j}\text{d}\uu^{j} & \approx \int \frac{\pi}{4} \left[1 - \left(\uu^{i}_{\alpha}\uu^{i}_{\beta} - \frac{\delta_{\alpha \beta}}{3}\right)\left(\uu^{j}_{\alpha}\uu^{j}_{\beta} - \frac{\delta_{\alpha \beta}}{3}\right) \right] \psi^{j}\text{d}\uu^{j} \\
& = \frac{\pi}{4} [1 - \uu^{i}\uu^{i}:\tens{Q}],
\end{split}
\end{equation}
where only terms up to second order have been kept. Finally, the full $\epsilon$ integral in \ref{unsimplified-anglej} is
\begin{equation} \label{eq:epsint}
    \int_{-\frac{L}{2}}^{\frac{L}{2}}{\text{d}\epsilon^{i}}\int_{-\frac{L}{2}}^{\frac{L}{2}}\text{d}\epsilon^{j} |\epsilon^{i}| (\epsilon^{i})^{2} = \frac{L^{5}}{32}.
\end{equation}
Inserting \ref{eq:epsint}, \ref{doi-tau} and \ref{4thclosed} into \ref{unsimplified-anglej} (and invoking the definition of $\langle \rangle_{j}$, Eq.~\ref{mean-field}) leads to 
\begin{equation}
    \left\langle  \frac{\left(\epsilon^{i} \right)^2}{|\vec{v}_{c}|} \right\rangle_{j} \approx \frac{3 \phi L^2 |\tens{D}|}{8D} \frac{[1 - \uu^{i}\uu^{i}:\tens{Q}]}{(\uu^{i} \cdot \nabla \vec{v} \cdot \uu^{i})^{2} + (\nabla \vec{v}:\tens{Q})^2}.
\end{equation}
This expression is not simple enough to use  directly, and we abstain from inserting it into Eq.~\ref{gen-Q-dot-with-fric} or Eq.~\ref{sig_main_fric}. In the specific case of simple shear this expression becomes
\begin{equation} \label{v-c-approx-appendix}
    \left\langle  \frac{\left(\epsilon^{i} \right)^2}{|\vec{v}_{c}|} \right\rangle_{j} \approx \frac{3 \phi L^2 }{8\sqrt{2}D\dot\gamma} \frac{[1 - \uu^{i}\uu^{i}:\tens{Q}]}{\left(\uu^{i}_x\uu^{i}_y\right)^2 + Q_{xy}^2}.
\end{equation}



\section{Derivation of numerical simulation} \label{app:numerical-deriv}  
We begin with Eq.~\ref{solidfric-omega-simu}, and using $g(\vec{\omega}_{\textrm{min}}) = \left \langle  \frac{(\epsilon^{i})^2}{|\vec{v}_{c}(\vec{\omega}_{\textrm{min}})|} \right\rangle_{j} $ and $\alpha = \Rhat \tilde{U} / \zeta_{r}$, consider the first order truncation
\begin{subequations}
\begin{align}
     \vec{\omega}^{i}_{\text{min}} & = \vec{\omega}_{0} - \alpha g(\vec{\omega}^{i}_{\text{min}}) \\
     & = \vec{\omega}_{0} - \alpha g(\vec{\omega}_{0} - \alpha g(\vec{\omega}^{i}_{\text{min}})) \\
     & \approx \vec{\omega}_{0} - \alpha g(\vec{\omega}_{0}).
\end{align}
\end{subequations}
This truncation is used alongside further approximations in Appendix~\ref{appendix:v-c-approx} to derive Eq..~\ref{v-c-approx-appendix}.

As $|\vec{F}_{N}|$, by equation \ref{press-def}, depends on the pressure and thus the stress, we must also treat the self-consistency introduced in Eq.~\ref{sig_main_fric}. For the case of only solid frictional contacts, we have
\begin{subequations}
    \begin{align}
    \boldsymbol{\sigma}_{\alpha \beta} &= \boldsymbol{\sigma}_{\alpha \beta}^{\textrm{F}} + \boldsymbol{\sigma}_{\alpha \beta}^{\textrm{DE}} \\
    & = 2 \bar{\rho} \mu_k|\vec{F}_{N}|\left\langle \left\langle \left[\frac{(\epsilon^{i})^2}{|\vec{v}_{c}|} \right] \left[\uu^{i}_{\alpha}\uu^{i}_{\beta}\,\left(\uu^{i}\!\cdot\!\vec{\nabla} \vec{v}\!\cdot\!\uu^{i}\right)
    +\vec{f} \cdot \frac{\partial \vec{f}}{\partial (\nabla v)_{\alpha \beta}}\right]  \right\rangle
    \right\rangle \\[5truept]
    &\qquad\qquad+ \bar{\rho} \int \left[ \zeta_{r} \vec{\Omega}_F \cdot \frac{\partial \vec{\Omega}_F}{\partial \nabla \vec{v}_{\alpha \beta}}  \right]  \psi  \text{d}\uu + \boldsymbol{\sigma}_{\alpha \beta}^{\textrm{DE}},
\end{align}
\end{subequations}
If we keep terms to first order in $\alpha$, this simplifies to
\begin{equation} \label{simu-simplified-stress}
    \boldsymbol{\sigma}_{\alpha \beta} \approx 2 \bar{\rho} \mu_k|\vec{F}_{N}|\left\langle \left\langle \left[\frac{(\epsilon^{i})^2}{|\vec{v}_{c}|} \right] \left[\uu^{i}_{\alpha}\uu^{i}_{\beta}\,\left(\uu^{i}\!\cdot\!\vec{\nabla} \vec{v}\!\cdot\!\uu^{i}\right) \right]  \right\rangle \right\rangle + \boldsymbol{\sigma}_{\alpha \beta}^{\textrm{DE}}.
\end{equation}
Along with Eq.~\ref{press-def} and Eq.~\ref{ac-def}, this allows us to express $|\vec{F}_{N}|$ directly:
\begin{subequations} \label{pre-arranged-fN}
\begin{align} 
    |\vec{F}_{N}| & = \frac{A_{c}\text{Tr}(\boldsymbol{\sigma})}{3}\\
    & \approx \left(\frac{\pi}{4 (1- S^2)} \right)^{\frac{2}{3}} \frac{D^2}{3\phi^{\frac{4}{3}}}\text{Tr}(\boldsymbol{\sigma}) \\
    & = \left(\frac{\pi}{4 (1- S^2)} \right)^{\frac{2}{3}} \frac{D^2}{3\phi^{\frac{4}{3}}} \left(2 \bar{\rho} \mu_k|\vec{F}_{N}|\text{Tr}(\tens{K}(\uu)) + \text{Tr}\left(\boldsymbol{\sigma}_{\alpha \beta}^{\textrm{DE}}\right) \right),
\end{align}
\end{subequations}
where we have inserted Eq.~\ref{ac-def} in the first line, and defined 
\begin{equation} \label{k-tens-def}
    \tens{K}_{\alpha \beta}(\uu) = \left\langle \left\langle \left[\frac{(\epsilon^{i})^2}{|\vec{v}_{c}|} \right] \left[\uu^{i}_{\alpha}\uu^{i}_{\beta}\,\left(\uu^{i}\!\cdot\!\vec{\nabla} \vec{v}\!\cdot\!\uu^{i}\right) \right]  \right\rangle \right\rangle.
\end{equation}

From Eq.~\ref{pre-arranged-fN}, we can solve for the form we require:
\begin{equation} \label{full-arranged-fn}
    |\vec{F}_{N}| = \frac{\left(\frac{\pi}{4 (1- S^2)} \right)^{\frac{2}{3}} \frac{D^2}{3\phi^{\frac{4}{3}}} \text{Tr}\left(\boldsymbol{\sigma}^{\textrm{DE}} \right)}{1 - 2 \bar{\rho} \mu_k \left(\frac{\pi}{4 (1 - S^2)} \right)^{\frac{2}{3}} \frac{D^2}{3\phi^{\frac{4}{3}}}\text{Tr}(\tens{K}(\uu)) }.
\end{equation}

From this equation we note an important condition: $|\vec{F}_{N}|$ cannot be a strictly decreasing function of $\text{Tr}(\tens{K})$. As our model introduces a self-consistency in $|\vec{F}_{N}| = |\vec{F}_{N}(\boldsymbol{\sigma})|$ and $\boldsymbol{\sigma} = \boldsymbol{\sigma}(|\vec{F}_{N}|)$, then the normal force must increase along with the stress, which is a linear function of $\tens{K}$ (see Eq.~\ref{simu-simplified-stress} and Eq.~\ref{k-tens-def}). From Eq.~\ref{k-tens-def} and Eq.~\ref{simple-shear-def}, we have 
\begin{equation} \label{trace-k}
    \text{Tr}(\tens{K}) = \dot{\gamma} \left\langle \left\langle \frac{(\epsilon^{i})^2}{|\vec{v}_{c}|} \right\rangle_{j} u_{x} u_{y} \right\rangle.
\end{equation}
As $\langle (\epsilon^{i})^2 / |\vec{v}_{c}| \rangle_{j}$ is positive (see Eq.~\ref{v-c-approx-appendix}), the condition is only violated in Eq.~\ref{full-arranged-fn} if $\langle u_{x}u_{y} \rangle< 0$ always. The nature of simple shear is to generally bring the orientation of rods to the extensional quadrant \cite{doiedwards}, \textit{i.e.} so $\langle u_{x}u_{y} \rangle > 0$, and the condition is satisfied.

The trace of the Doi-Edwards stress in simple shear is (using Eq.~\ref{simplified-de-stress} and Eq.~\ref{simple-shear-def})
\begin{equation} \label{tr-de-stress}
    \text{Tr}\left(\boldsymbol{\sigma}^{DE}\right) = \overline{\rho} \frac{\zeta_r}{2} \dot{\gamma} |Q_{xy}|.
\end{equation}
Note that in the above we have made one artificial adjustment to the expression, namely, introducing the absolute value function to the $Q_{xy}$ term. As our simulation is stochastic, fluctuations in the orientations of particles can cause the sign of $Q_{xy}$ to switch. Allowing $\text{Tr}\left(\boldsymbol{\sigma}^{DE}\right) < 0$ is unphysical -- $|\vec{F}_{N}|$ could then be negative -- and does not represent a compressive stress squeezing particles together as intended in order to initiate contacts \cite{gillissen2020constitutive}.

Inserting Eq.~\ref{tr-de-stress} in Eq.~\ref{full-arranged-fn} and using the result along with Eq.~\ref{v-c-approx-appendix} in Eq.~\ref{solidfric-omega-simu}, we have
\begin{subequations}\label{app-w-eq-solve}
\begin{align}
\vec{\omega}^{i}_{\text{min}} & =  \vec{\omega}_{0} - \frac{\Rhat\tilde{U}}{\zeta_{r} + \mu_{k} \frac{\left(\frac{\pi}{4 (1- S^2)} \right)^{\frac{2}{3}} \frac{D^2}{3\phi^{\frac{4}{3}}} \overline{\rho} \frac{\zeta_r}{2} \dot{\gamma} |Q_{xy}|}{1 - 2 \bar{\rho} \mu_k \left(\frac{\pi}{4 (1 - S^2)} \right)^{\frac{2}{3}} \frac{D^2}{3\phi^{\frac{4}{3}}}\text{Tr}(\tens{K}(\uu)) } \frac{3 \phi L^2 }{8\sqrt{2}D\dot\gamma} \frac{[1 - \uu^{i}\uu^{i}:\tens{Q}]}{\left(\uu^{i}_x\uu^{i}_y\right)^2 + Q_{xy}^2} }   \\
& = \vec{\omega}_{0} - D_{r}\frac{\Rhat\tilde{U} / k_{B}T}{1 + \mu_{k} \frac{\left(\frac{\pi}{4 (1- S^2)} \right)^{\frac{2}{3}} \frac{2 |Q_{xy}|}{3 \pi D L \phi^{\frac{1}{3}}} }{1 - \mu_k \left(\frac{\pi}{4 (1 - S^2)} \right)^{\frac{2}{3}} \frac{8}{3 \pi L \phi^{\frac{1}{3}}} \text{Tr}(\tens{K}(\uu)) } \frac{3 \phi L^2 }{8\sqrt{2}}\frac{[1 - \uu^{i}\uu^{i}:\tens{Q}]}{\left(\uu^{i}_x\uu^{i}_y\right)^2 + Q_{xy}^2} } \\
& \equiv  \vec{\omega}_{0} - D_{r}\frac{\Rhat\tilde{U} / k_{B}T}{1 + C(\phi, L/D, \mu_{k})\frac{[1 - \uu^{i}\uu^{i}:\tens{Q}]}{\left(\uu^{i}_x\uu^{i}_y\right)^2 + Q_{xy}^2} } ,
\end{align}
\end{subequations}
where we have introduced the scalar function $C$ for simplicity, used $\bar{\rho} = \frac{\phi}{\frac{\pi}{4}D^{2}L}$ in the second line, and $D_{r} = \frac{k_{B}T}{\zeta_{r}}$ in the third line. 

\color{black}
\end{widetext}
\clearpage

\bibliography{proposal}

\end{document}